\documentclass[twocolumn,aps,prb]{revtex4}
\usepackage{graphics,graphicx}
\usepackage{dcolumn}
\usepackage{bm}

\def\eb{\begin{equation}}   
\def\ee{\end{equation}}     
\def\ea#1{\begin{eqnarray} #1 \end{eqnarray}}   

\def\shro{Schr\"odinger}

\def\s{\sin\theta}

\def\G{\hat{G}}

\def\T{\hat{T}}

\def\grad{\vec\nabla}

\def\ra{\rightarrow}

\def\t{\tilde}

\def\of#1{\left(#1\right)}

\def\prt#1{{\partial \over \partial #1}}
\def\prtsq#1{{\partial^2 \over \partial {#1}^2}}

\def\eq#1{Eq.~(\ref{#1})}
\def\eqs#1#2{Eqs.~(\ref{#1}) and (\ref{#2})}

\def\sof#1{\left[ {#1} \right]}

\def\P{\hat{P}}

\def\prt#1#2{{\partial #1 \over \partial #2}}
\def\prtsq#1#2{{\partial^2 #1 \over \partial {#2}^2}}

\def\a{\alpha}
\def\b{\beta}
\def\g{\gamma}
\def\G{\Gamma}
\def\d{\delta}
\def\k{\kappa}
\def\s{\sigma}
\def\l{\lambda}
\def\t{\tau}
\def\T{\mathcal{T}}
\def\L{\mathcal{L}}
\def\p{\partial}
\def\P{\nabla}

\begin{document}

\title{Trajectory-based Theory of Relativistic Quantum Particles}

\author{Bill Poirier}
\affiliation{Department of Chemistry and Biochemistry, and
         Department of Physics, \\
          Texas Tech University, Box 41061,
         Lubbock, Texas 79409-1061}
\email{Bill.Poirier@ttu.edu}

\begin{abstract}

Recently, a self-contained trajectory-based formulation of non-relativistic quantum
mechanics was developed [Ann. Phys. {\bf 315}, 505 (2005); Chem. Phys.
{\bf 370}, 4 (2010); J. Chem. Phys. {\bf 136}, 031102 (2012)], that makes no use of 
wavefunctions or complex amplitudes of any kind. Quantum states are represented as 
ensembles of real-valued quantum trajectories that extremize a suitable action. Here,
the trajectory-based approach is developed into  a viable, generally covariant,  
relativistic quantum theory for single (spin-zero, massive) particles.  Central to this development 
is the introduction of a new notion of global simultaneity for accelerated particles---together with 
basic postulates concerning probability conservation and causality.  The latter postulate 
is found to be violated by the Klein-Gordon  equation, leading to its well-known problems as a 
single-particle theory. Various examples are considered, including the time evolution of a 
relativistic Gaussian wavepacket.

\end{abstract}

\maketitle                 


\section{INTRODUCTION}
\label{intro}

In this document, we develop a new formulation of  single-particle relativistic 
quantum mechanics.  Traditionally,  the formulation of quantum 
mechanics proceeds via a set of 
postulates,\cite{vonneumann,cohen-tannoudji,bohm,mcquarrie,messiah} which we 
do not find it necessary to repeat here. We do note, however,  that the order, 
precise content, and even total number of quantum postulates, vary from one treatment 
to the next.  This situation might be taken as an indication of the controversy or 
uncertainty that still exists---particularly around those postulates having to do 
with quantum  measurement.  The latter are especially nettlesome in the context of
relativistic quantum mechanics---where, e.g.,  it may not be entirely clear how to 
reconcile the ``instantaneous'' collapse of the wavefunction with subluminal
propagation. On the other hand,  the traditional quantum treatments all
do agree on the first and  most important postulate---that the state of a system 
be completely described by the quantum wavefunction, $\Psi$. 

The wavefunction has thus  always enjoyed a hallowed status in quantum mechanical 
theories, despite much historical and ongoing disagreement about its precise 
interpretation or physical significance.\cite{styer02,madelung26,bohm52a,bohm52b,holland,wyatt,durr92,berndl95,einstein35,ballentine70,home92,everett57,dewitt70,wheeler}  Even those ``alternative'' interpretations
of quantum mechanics that dare to challenge the first postulate still tend to respect the 
supremacy of $\Psi$. An example would be Bohmian 
mechanics,\cite{madelung26,bohm52a,bohm52b,holland,wyatt,durr92,berndl95}  
which adopts a hybrid
ontology wherein it is the wavefunction {\em plus} a quantum trajectory,
together, that  are needed to completely specify the quantum state.
For many physicists, it may be difficult to even  conceive of a quantum 
theory that makes no direct or indirect recourse to wavefunctions---after
all, $\Psi$, appears in every one of the five (or six) standard postulates.  
Nevertheless, exactly just such a theory was  recently formulated for non-relativistic 
quantum mechanics.\cite{bouda03,holland05,poirier10nowave,holland10,poirier11nowaveCCP6,poirier12nowaveJCP,poirier12ODE}

For a number of reasons, it makes sense to try to extend the previous work to the
{\em relativistic} case.  As presented in this document, this goal is now also
achieved---at least in the context of a single, spin-zero, 
massive, relativistic quantum particle, propagating on a flat Minkowski spacetime, 
with no external fields.  Such a system might represent, e.g., a single Higgs boson 
particle---apropos to which, the recent news from CERN\cite{CERN-higgs} is serving to 
stimulate demand for new and fresh approaches. In any case, generalizations for curved 
spacetimes, external fields, photons, particles with spin, multiple particles,  etc., 
together with  detailed analyses of  symmetry and stability properties, are planned 
for the future (although for each such development, a varying degree of required 
effort is anticipated).

To be clear, the present approach is {\em not} a quantum field theory (QFT), but 
rather, a conserved particle approach---in that sense, like the non-relativistic time-dependent 
\shro\ equation (TDSE). As a fundamental theory of relativistic quantum mechanics, 
it is safe to say that a particle-based strategy has been largely abandoned for many
 decades (with some notable exceptions\cite{bialynicki94,sipe95,rosenau11}).
 Of course, the reasons for this date back to the earliest attempts to ``relativize'' the TDSE, 
 starting with the Klein-Gordon (KG) equation in 
1926.\cite{bohm,messiah,holland,debroglie56,feshbach58,carroll,aharonov69,blokhintsev,ranada80,kyprianidis85}  
Whereas the TDSE is first order in time and second order in
space, the KG equation (which treats space and time on an equal footing) is second order
in both---wherein lies the source of most of its difficulties.  In particular, this leads to:  
(1) negative energy solutions, as well as (2) negative (indefinite) probability 
densities.\cite{bohm,messiah,holland,aharonov69,blokhintsev,ranada80,kyprianidis85}

In 1928,  Dirac improved matters somewhat, with the development of his famous
first order (in both time and space) but multi-component equation, describing 
spin $1/2$ particles.\cite{messiah,holland,dirac28a,dirac28b}  Dirac effectively solved 
problem (2), but not problem (1).   
By 1934 however, the ``real'' solution to both problems was hit upon---i.e., second 
quantization, and the development of QFT.\cite{messiah,thirring,bogoliubov,schweber}
Though obviously serving us well in the ensuing decades, particularly for processes
involving the creation/annihilation of particles, it can be argued that QFT introduces as
many problems as it solves (e.g., pertaining to causality\cite{bogoliubov,schweber} 
as well as renormalization), and in any event greatly
complicates matters, both theoretically and conceptually. Presumably, a 
viable, rigorous, single-particle theory of relativistic quantum mechanics would 
therefore still be welcomed with open arms.

The time is now ripe to revisit this notion. What makes us imagine that we can 
succeed where Klein, Gordon, and Dirac evidently failed? The crucial development is 
the recent wavefunction-free reformulation of non-relativistic quantum mechanics, 
alluded to above.\cite{bouda03,holland05,poirier10nowave,holland10,poirier11nowaveCCP6,poirier12nowaveJCP,poirier12ODE}
This approach is trajectory based, and in that sense reminiscent of Bohmian mechanics.
Unlike the Bohm theory, however, here,  the traditional wavefunction, $\Psi(t,{\bf x})$, is entirely 
done away with, in favor of the {\em trajectory ensemble}, ${\bf x}(t,{\bf C})$ (where ${\bf C}$ 
labels individual trajectories) as the fundamental representation of a quantum state.  
The ensemble is such that there is exactly one trajectory passing through every point in space
($\bf x$) at any given point in time ($t$). 
The ensemble satisfies its own partial differential equation (PDE) 
that is mathematically equivalent to the TDSE---even though it is nonlinear, real valued, 
second order in $t$ and fourth order in ${\bf C}$.  It also satisfies an action principle, for which 
the Lagrangian consists of the usual classical part, plus a quantum contribution that incorporates  
{\em intertrajectory interactions} (i.e., partial derivatives of ${\bf x}$ with respect to ${\bf C}$).  

From the perspective of developing a relativistic generalization of quantum theory, 
the trajectory-based approach is extremely well suited to the task.  
As in standard relativity theory (and non-relativistic classical mechanics), 
quantum solution trajectories are obtained as those that extremize the appropriate quantum 
action quantity---whose relativistic form may be guessed as the Lorentz-invariant (or generally 
covariant) version of the non-relativistic quantum action described above.  
As in relativity theory, also,  the quantum solution trajectory PDEs are 
inherently nonlinear.  Although in the non-relativistic case, these happen to be
equivalent to a linear wave PDE (i.e., the TDSE for $\Psi$), there is no 
 {\em a priori} reason to expect such a relationship to hold in the relativistic case. 
 Indeed, if a viable (i.e., non-KG)  relativistic linear wave PDE describing individual
 quantum particles were possible,  then it probably would have been discovered 
 decades ago by one of the aforementioned luminaries...
 
 At any rate, in this work we develop a  relativistic  generalization of our earlier
  non-relativistic trajectory-ensemble-based theory for quantum particles.  
 It must be emphasized that {\em new physics is being predicted here}---which, 
 in principle, could be validated or refuted by comparison with experiment. 
 Because we are operating largely in uncharted waters, it is possible that the present 
 form of our equations may have to be modified (as was famously the case with 
 Einstein's own equations); at the same time, however, 
 general covariance considerations {\em greatly} restrict the form that such alternate  
 dynamical laws might take. In any event, the form presented here is likely the
 simplest and most reasonable.  We note that although only special relativity (SR)
 {\em per se} is considered here, the mathematical development of our approach
 relies heavily on curvilinear coordinate systems---and therefore, on the framework
 and tools of general relativity (GR).\cite{carroll,weinberg}
 
 Central to our approach is our (evidently) new definition of {\em simultaneity for 
 accelerated particles.}  Even in SR, there is no good notion of the set of all 
 spacetime events that occur simultaneously (from the particle's perspective) 
 with a given event on the particle's worldline, if it is accelerating. In this context, 
 simultaneity is sometimes defined in the usual unaccelerated manner---i.e., 
 using local  inertial or comoving frames.\cite{carroll,weinberg,rindler}  
 This strategy fundamentally fails, however, because it predicts multiple 
 reoccurrences of the same spacetime events  (e.g. pivot points\cite{rindler,dolby03}), 
 as well as the incorrect time ordering for distant,  timelike separated events. 
  Quite remarkably, the present, trajectory  ensemble approach allows for a natural
  and rigorous generalization of the simultaneity concept for an arbitrarily-moving
  particle---essentially,  because the quantum nature of the particle imparts a global 
  character to it.
   
 The new relativistic quantum trajectory PDEs can be analyzed
 in various ways.  In inertial coordinates,
 one obtains a form that is  similar to the KG equation---yet 
 differs in one very crucial respect (discussed in Sec.~\ref{KG}). In retrospect,
 from a trajectory ensemble vantage, one can see clearly exactly where 
KG ``got the physics wrong,''  in their attempt to relativize the TDSE.  
However, in order to do so, one must transform from the inertial frame
to a certain curvilinear (albeit naturally arising) coordinate system, in terms of
which the new PDEs are  fourth order in space, and only 
{\em second} order in  time (i.e., just like the non-relativistic quantum 
trajectory PDEs). The ramifications of this---and more generally, 
of the apparent inherent nonlinearity of the new PDEs---are not yet 
entirely clear. Thus,  it may turn out that the present form
 is not always stable (or that other unanticipated problems may eventually 
 manifest)---although instability has not yet been observed, e.g., in the 
 examples considered in Sec.~\ref{examples}.  However, if such difficulties were to arise in the future,
 the author's view is that it should serve as a call to arms to look for a suitable rectifying 
 modification of the present formulation---rather than as a condemnation of the 
 general approach presented here, which seems to have much to offer. 

This document is quite long and comprehensive, as the requisite formal development 
is rather involved. We thus provide here a detailed overview of the remaining sections, 
with a specification of those subsections that might be skipped on an initial reading.
Sec.~\ref{basic} addresses the basic mathematical structures that underpin the trajectory-based
approach, in the relativistic context of a Lorentz-invariant four-dimensional (4d) spacetime.
Sec.~\ref{preliminaries} mainly establishes the notation and conventions as adopted here, but also 
introduces the trajectory ensemble four-velocity field; much of this material is standard,
and can be skipped by one versed in SR theory. The all-important simultaneity 
submanifold is then promptly constructed in Sec.~\ref{simultaneity}.  In Sec.~\ref{ensemble-time}
, the ``ensemble time'' parameter is introduced, as a  label for the different simultaneity submanifolds; 
this is found to be incompatible with the usual proper time, for reasons related to the 
famous twin ``paradox.''  

Sec.~\ref{general-prob} introduces the natural curvilinear coordinate
system alluded to above (Sec.~\ref{general}), together with various probability density quantities.
The most relevant equations in Sec.~\ref{general} are Eqs.~(\ref{natcoord}), (\ref{gblock}), 
and~(\ref{g00}), where the last two describe the form of the metric tensor in
natural coordinates. Sec.~\ref{prob3d} introduces the first postulate of the trajectory-based
approach (pertaining to probability conservation) and discusses the spatial scalar probability density, 
whereas Sec.~\ref{prob4d} considers the scalar invariant, 4d, and flux four-vector generalizations.   
Sec.~\ref{cov-continuity} derives the covariant continuity equation,
and Sec.~\ref{uniformizing} discusses a particularly useful set of natural coordinates. The last three
subsections of Sec.~\ref{general-prob} are not as critical for an initial reading.

Having laid out much of the mathematical framework in Secs.~\ref{basic} and~\ref{general-prob}, 
Sec.~\ref{dynamical}  addresses
dynamical considerations. The early part of Sec.~\ref{classical} is critical, as it introduces the
second postulate of the trajectory-based approach, pertaining to causality. This sensible 
condition is satisfied by standard classical and non-relativistic quantum mechanics (TDSE), but 
not---it turns out---by the relativistic KG equation. Sec.~\ref{electromagnetism} is a somewhat technical 
exposition on time reparametrization that may be skipped on an initial reading, whereas
Sec.~\ref{nonrel} is a review of the previous non-relativistic trajectory-based formulation, couched in 
the covariant language of GR.   

The meat of the new theory is presented in
Sec.~\ref{relativistic}.   The new relativistic quantum PDE itself is readily 
obtained in Sec.~\ref{relPDE} [\eqs{reldyn}{relfQ}], although not in a form that is practically useful.   
That this  PDE satisfies the principle of least action is established in Sec.~\ref{relativistic-action}, 
for those who wish to see how this comes about. 
The final, more practical form of the PDE  [\eq{finalPDE}] 
is then derived in Sec.~\ref{conversion}---in which, also, an unexpected connection is established
between the quantum and gravitational potentials.  Sec.~\ref{limits} considers various limiting
forms of the PDE (e.g., the non-relativistic limit), and Sec.~\ref{KG} presents a detailed comparison 
with KG theory; both may be skipped on an initial reading.  Various examples are 
presented in Sec.~\ref{examples}, with the relativistic Gaussian wavepacket of Sec.~\ref{gaussian}
 the most relevant. Finally, a concluding discussion is provided in Sec.~\ref{conclusion}.


\section{Basic Mathematical Structure}
\label{basic}

\subsection{Preliminaries}
\label{preliminaries}

Let $M$ be a 4d Reimannian manifold, representing the
spacetime of a single, spin-zero, relativistic quantum particle of mass $m$.
For purposes of this study, $M$ is presumed flat (Ricci scalar $R\! = \!0$).
A global inertial frame can therefore be defined, in terms of which the inertial
coordinates are $x^\a = (c\,t, {\bf x}) = (c\,t, x^l) =  (c\,t, x, y, z)$, and the metric
tensor $\eta_{\a\b}$ is the usual Minkowski one,
\eb
	\tilde\eta = \left( \matrix{ -1 & 0 & 0 & 0 \cr
	                                      0 & 1 & 0 & 0 \cr
	                                      0 & 0 & 1 & 0 \cr
	                                      0 & 0 & 0 & 1 } \right).
\ee  
Note that we adopt the -+++ convention for the metric signature; also,
factors of $c$ are always explicitly indicated.  Thus, the line element 
$ds^2 = \eta_{\a\b}\, dx^\a dx^\b$ has units of length squared, whereas the 
proper time, $\t$, defined via 
\eb
	d\t^2 = -  {1 \over c^2}\, \eta_{\a\b}\, dx^\a dx^\b,
	\label{dtaudef}
\ee
has units of time.  The Greek indices $\a$, $\b$, $\g$,  $\d$, run over
the four spacetime inertial coordinate labels, i.e. 0,1,2,3, whereas 
$\mu$, $\nu$, $\k$, $\s$ serve a similar function for general curvilinear
coordinate systems (denoted $X^\mu$).  
Latin indices run over spatial (or spacelike) coordinate labels 1,2,3, as per usual,
with $l$, $m$ (not to be confused with mass) used for inertial
coordinates, and $i$, $j$, and $k$ for general coordinates.

A {\em path ensemble} (candidate solution trajectory ensemble) is uniquely
specified via a contravariant vector field $U^\a$ (in inertial coordinates), satisfying
\eb
	\eta_{\a\b}\, U^\a U^\b = -  c^2  \label{fourvcond}
\ee 
everywhere. Equation~(\ref{fourvcond}) 
above implies that $U^\a$ is everywhere timelike, and may be interpreted 
as a four-velocity field, i.e. 
\eb
	U^\a = {d x^\a \over d \t}. \label{fourv}
\ee
Integration of \eq{fourv} then gives rise to a set of 1d submanifolds that 
foliate $M$, and therefore do not cross (even self-intersections are
prohibited by the topology; the submanifolds are inextendible).  This
family of timelike curves thus constitutes the ensemble of paths, 
or candidate solution trajectories.

There is exactly one path passing through every point $p$ in $M$;  also, a 
one-to-one correspondence exists between paths (whose codimension is 3) 
and spatial points (the set ${\bf x}$ for fixed $t$). 
Moreover,  all of the above properties are preserved under Lorentz
transformations of the inertial frame, i.e.
\eb
	x^\a \ra x^{\a'} = {\Lambda^{\a'}}_{\!\a}  \, x^\a.
	\label{LT}
\ee
Specifically the transformed four-velocity field, $U^{\a'}$, satisfies primed versions of
 \eqs{fourvcond}{fourv}, and a correspondence can be established between paths and 
${\bf x}'$ points,  for fixed $t'$. 

We next introduce a set of {\em path labeling parameters},
${\bf C} = C^i$, which uniquely identify individual paths within the ensemble. The $C^i$ are
not yet coordinates per se, although later we will construct curvilinear coordinate systems
from them.  Note that the $C^i$ values do not change under Lorentz transformations. 
Since the labeling parameters can (but in general will not) be 
identified with ${\bf x}$ at some specific $t$ (for some specific inertial frame), 
we regard the $C^i$ as spacelike parameters.  In any event, for a given path 
ensemble and choice of inertial frame, ${\bf x}(t, {\bf C})$ and ${\bf C}(t, {\bf x})$ 
are well-defined inverse diffeomorphisms of each other, for fixed values of  $t$.

Note that the above claims are subject to certain caveats, such as the possible
existence of a measure-zero set of exception points.\cite{poirier10nowave}

\subsection{Simultaneity submanifolds}
\label{simultaneity}

Consider the tangent space $T_p$ for a point $p$ in $M$.  A 3d spacelike 
``orthogonal subspace'' can be defined as the set of all vectors $W^\b$ in $T_p$ 
that are orthogonal to $U^\a$:
\eb
	\eta_{\a\b}\, U^\a\, W^\b = 0  \label{orthosub}
\ee 
The orthogonal subspace is a linear vector space in its own right; the collection
of all such subspaces for every point $p$ in  $M$ forms a subbundle.  
We presume that the orthogonal subspaces can be integrated outward 
from the point $p$, to construct a corresponding integral submanifold.  
Specifically, this is a 3d embedded submanifold of $M$, a spacelike 
hypersurface, that intersects every path exactly once. 
By starting this procedure from each point $p$ that lies along 
a particular reference path,  a one-parameter family of such hypersurfaces may be 
constructed, which are non-intersecting, and otherwise foliate $M$.   By construction,
at every point $p$ in $M$, the four-velocity vector $U^\a$ is normal to the
hypersurface passing through that point.

We hereby refer to the above 3d hypersurfaces as ``simultaneity submanifolds.''
This terminology is justified through the following physical arguments.  Consider
a particle on a worldline passing through event $p$. At that instant, the velocity
four-vector $U^\a$ defines the local forward time direction for that particle. Likewise,
the orthogonal subspace defines the local spacelike directions for that particle---that
is to say, the local set of events that occur simultaneously with $p$, from the particle's
perspective.  This much, at least, is consistent with the idea of local inertial frames, 
and more importantly, with the Einstein Equivalence Principle.\cite{carroll,weinberg}  
The problem in standard relativity theory, of course, is that of extending these {\em local} 
notions of simultaneity in a {\em global} manner.  Using inertial coordinate frames, 
this can be achieved for the special case of unaccelerated motion, but it is problematic for
accelerating particles (a limitation that in hindsight, should perhaps seem a bit odd).

In any event, the quantum relativistic theory developed here provides a 
{\em global concept of simultaneous events} for a single, arbitrarily-moving particle, 
in the form of the simultaneity submanifolds described above.  This is perhaps
most physically meaningful if one adopts a ``many worlds''-type ontological interpretation
of the multiple particle paths/trajectories, according to which each trajectory worldline literally
represents a different world, as has been discussed in previous 
work.\cite{poirier10nowave,poirier11nowaveCCP6,poirier12nowaveJCP}
The one particle is thus comprised of many ``copies,'' distributed across all space.  Locally,
the structure of the orthogonal subspace described above ensures that each particle copy 
agrees with its nearest neighbors as to which events occur simultaneously.  Because
of the global distribution of copies, however, this notion can be extended globally throughout 
all of space. In this fashion---and much like Einstein's own
orthogonal-ruler-and-clock construction of inertial frames---one builds a  global manifold of 
simultaneous events,  agreed upon by all particle copies, regardless of whether 
some or all of those copies are accelerating, and despite the fact that they never cross
paths.

Note that the simultaneity submanifolds are {\em not}  absolute, in the sense of being 
agreed upon by  all {\em observers}.   A different quantum observer (or particle) would
 have its own copies,  its own trajectory ensemble, and (in general) an 
entirely different set of simultaneity submanifolds. This is as it should be. Note that for
two quantum observers to agree completely on simultaneity, every trajectory in one 
ensemble must match the corresponding trajectory in the other ensemble (i.e., the 
two trajectory ensembles must be identical).   

\begin{figure}
\includegraphics[scale=0.65]{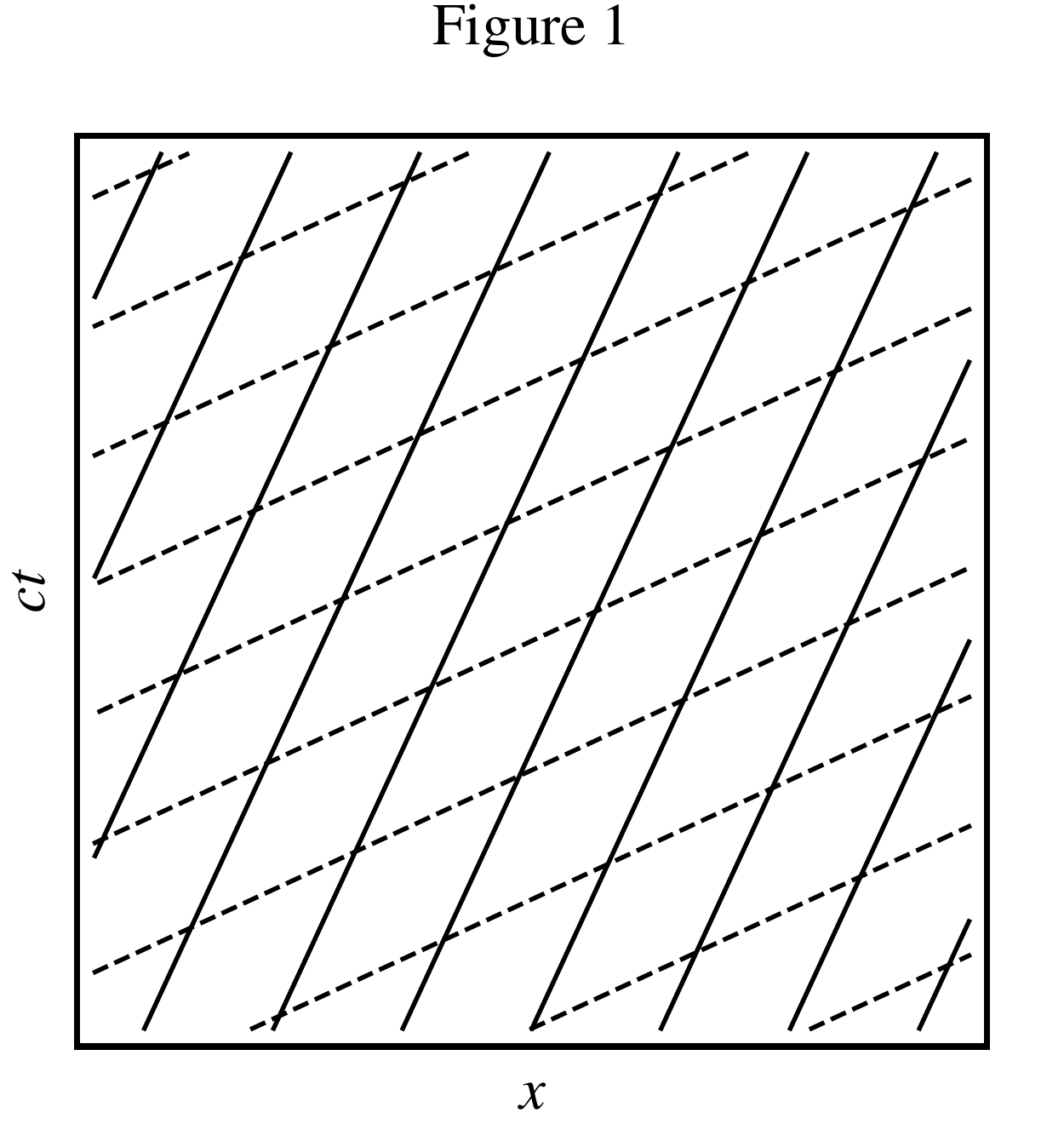}
        \caption{Contour plot of 1d inertial coordinates, $x^{\a'}=(c\, t', x')$,
as functions of the inertial coordinates,  $x^\a=(c\,t,x)$,
to which they are related via a Lorentz transformation [\eq{LT}].  The dashed
lines are the contours of $t'(c \,t, x)$. These represent ``simultaneity 
submanifolds''---i.e., sets of spacetime events that occur 
simultaneously---for an inertial observer moving with the primed frame.
The solid lines are the contours of $x'(c\, t, x)$, 
any one of which could represent the trajectory of the moving
inertial observer or particle, according to standard SR theory.  Collectively, the
solid lines constitute a {\em relativistic quantum trajectory ensemble},
describing a single relativistic particle undergoing quantum inertial motion.}
        \label{fig-inertial}
\end{figure}

Likewise,  a  ``quantum inertial observer'' is characterized via an ensemble of 
parallel straight-line trajectories---i.e., by
\eb
	U^\a = \rm{const}^\a.
	\label{inertialdef}
\ee  
Thus,  for example, if $x^{\a'}$ satisfies \eq{LT}, then the contours of $x^{0'}$ (expressed 
in the $x^\a$ coordinate system) define the simultaneity submanifolds for a quantum inertial 
particle, whose trajectory orbits are given by the intersections of the $x^{l'}$ contours.
As indicated in Fig.~\ref{fig-inertial},  this special case is of course, entirely consistent with the usual
global notion of simultaneity for unaccelerated particles---except that here, it is obtained 
in a more rigorous, essentially completely local fashion. The reason is that 
$U^\a$ is a four-velocity vector field---defined on all of $M$, rather than just along a
single trajectory.

More generally---i.e., for accelerated motion---the relativistic quantum trajectories and 
simultaneity submanifolds behave more along the lines indicated in Fig.~\ref{fig-Gaussian}.  
Unlike comoving frames,\cite{carroll,weinberg,rindler}  the simultaneity submanifolds of the 
present theory are curved.
This is appropriate, given that the trajectories are also curved (albeit in the extrinsic
rather than intrinsic sense).  The curvature of the
simultaneity submanifolds also enables them to avoid intersecting each other---thus
also avoiding the problems of  multiple reoccurences and incorrect time orderings that 
plague the comoving frame approach.  Finally, the simultaneity submanifolds are
everywhere orthogonal to the trajectories, and thus each submanifold  consists only 
of spacelike separated events.

\subsection{Ensemble time, and the generalized twin ``paradox''}
\label{ensemble-time}

It should be noted that the ability to construct global simultaneity submanifolds is not
automatic, but in fact, induces a slight constraint on the allowed form of $U^\a$. Using
Frobenius' theorem,\cite{carroll,schutz}  and the fact that  the $U^\a$ field is (presumed) smooth, 
it can be shown that only fields for which $h\, U^\a$ is four-curl-free (for some scalar 
field $h$) are permissible. (Technically, this is slightly more restrictive than the 
actual condition, but it is a sufficient condition that suits our purposes better).   
This condition  implies that  the four-velocity field is the normalized gradient of 
some scalar field $\l$:
\eb
U^\a = c {-\eta^{\a \b} \, \p_\b \l \over \sqrt{-\eta^{\g\d}\, \p_\g \l \,\, \p_\d \l}},
\label{lambdaeq}
\ee
where $\eta^{\a\b}=\eta_{\a\b}$ and $\p_\a = \p / \p x^\a$ (and it is presumed that 
$\p_0 \l > 0$).

The quantity $\l$ is a timelike parameter that we call an ``ensemble time.''
Actually, it is a full-fledged scalar field, and can therefore be interpreted 
as a global time {\em coordinate}.   \eqs{orthosub}{lambdaeq} imply that the 
 $\l = \rm{const}$ {\em contours are the simultaneity submanifolds}. 
The term ``ensemble time'' is therefore justified, as all members of the ensemble
(i.e., all particle copies) agree that events corresponding to the same value of $\l$
occur simultaneously.  The actual $\l$ value itself, however, is not uniquely defined. 
In general, any transformation of the form $\l \ra {\l}' =  {\l}'(\l)$ yields a new ensemble
time coordinate with the same contours, and which otherwise also satisfies \eq{lambdaeq}.  
For the moment, we treat all such choices equally. Later, we will 
identify special candidates to serve as the ``ensemble proper time,'' $\l=\T$ 
(Secs.~\ref{uniformizing} and~\ref{conversion})---i.e., the ensemble analog of the usual single-path 
proper time, $\tau$. 

Even for a path ensemble, it is straightforward---and often convenient---to 
construct a true proper time coordinate, $\t$, as a scalar field on $M$.   For each path 
in the ensemble,  labeled by ${\bf C}$, one simply chooses a reference point $p$ at 
which $\t$ is taken to be zero, and  then integrates \eq{dtaudef} along the path to find the value 
of $\t$ at all other points.  There is thus a freedom in the definition of the $\t$ coordinate,
associated with the particular choice of reference point $p$ for each path. 
This freedom corresponds to coordinate transformations of the form 
\eb
	\t \ra \t'  = \t' (\t, {\bf C}) = \t + \Delta({\bf C}),
	\label{taudef}
\ee
where the shift, $\Delta({\bf C})$, varies across paths.  

Although the $\tau$ coordinate is useful in its own right, note that any choice of $\tau$
is in general incompatible with any $\l$ coordinate of the form described above.
In particular, for two different paths of a given ensemble, both starting at the same simultaneity 
submanifold ($\l$ contour) and ending at another, the elapsed ensemble time $\Delta \l$ is
the same, but the elapsed proper times $\Delta \t$ are (generally) different.  Thus, 
$\tau$ per se cannot generally serve as a good ensemble time coordinate, $\l$.

The situation above is not problematic, and in fact, reflects nothing more than a
generalized version of the well-known (but poorly named) twin ``paradox.'' 
The conventional twin paradox has one twin leaving the other at spacetime point
$p_0$, only to rejoin him or her at a later point $p_f$, after having undergone accelerated 
motion. The accelerated trajectory of the second twin is necessary, in order that their
paths may recross (the stay-at-home twin is presumed to undergo 
inertial motion).  Since the starting point for both twins is in fact the same event $p_0$,
there can be no question but that the departures occur simultaneously. Likewise, the 
reunion  at $p_f$ is the same event for both twins, and must therefore also occur 
simultaneously. Nevertheless, we know that the elapsed proper times for the two 
twins are different, with the stay-at-home twin having aged (in some renditions, 
very significantly) more than his or her more adventurous sibling. 

A similar situation characterizes our ensemble of paths---except that we
now have a global definition of simultaneity, so that we can specify that two 
such paths begin ``at the same moment'' even if the initial spacetime events are
different. Likewise, we can uniquely specify the simultaneous ``end'' of 
the two paths, as the points where these paths intersect a different $\l$  contour. 
The global property of the simultaneity submanifolds  is indeed required, 
as any two paths within a given path ensemble never cross.    
Even though the two paths start and end simultaneously, the generalized
twin paradox implies that there is no reason to expect the two elapsed $\Delta \t$ values
to be the same---and indeed they are not, in general.   
For the special case where all paths are moving 
inertially---i.e., for a ``quantum inertial observer'' (Fig.~\ref{fig-inertial})---then the elapsed proper times 
for all paths are equivalent,  and it is permissible in this instance to 
use $\t$ as an ensemble time coordinate, $\l$.   We thus make it
a requirement of any reasonable definition for an ensemble proper time, $\T$,
that it should reduce to the usual proper time, $\t$, in  the special case of a quantum
inertial observer.


\section{General Coordinates and Probability}
\label{general-prob}

\subsection{General (curvilinear) coordinates and natural coordinates}
\label{general}

All of the equations of Sec.~\ref{basic} have been presented in a way that is
manifestly covariant, with respect to arbitrary coordinate transformations
(diffeomorphisms). The general (or curvilinear) coordinate version of 
all such results is obtained by replacing the indices 
$\a$, $\b$, etc. with $\mu$, $\nu$, etc., the inertial coordinates $x^\a$ 
with general coordinates $X^\mu$,  and the Minkowski metric tensor 
with the generalized form, 
\eb
	g_{\mu\nu}   =  \eta_{\a\b}\, \prt{x^\a}{X^\mu}\,\prt{x^\b}{X^\nu} 
         \label{gdef}
\ee
Note also that in principle, all partial derivatives $\p_\a$ that appear in 
Sec.~\ref{basic} should be replaced with covariant derivatives,\cite{carroll,weinberg} 
denoted here as $\P_\mu$.
However, these appear only in \eq{lambdaeq}, where they are applied
to a scalar invariant field ($\l$), for which it is well-known that 
$\P_\mu = \p_\mu$ (in any coordinate system).

It is sometimes convenient to write \eq{gdef} in matrix form,
\eb
	\tilde g   =  {\tilde J}^T \cdot \tilde \eta \cdot \tilde J,
	\label{gdefmat}
\ee
where ${J^{\a}}_{\!\mu} = \p_\mu x^\a$ is the Jacobian matrix for the 
coordinate transformation $x^\a \ra X^\mu$.  We also 
define the inverse transform Jacobian matrix, $\tilde K$, as 
${K ^{\mu}}_{\!\a} = \p_\a X^\mu$.  Being true inverses of each other, 
${J ^{\a}}_{\!\mu} \,{K ^{\mu}}_{\!\b} = \d^\a_\b$,  and 
${K ^{\mu}}_{\!\a}\,{J ^{\a}}_{\!\nu} = \d^\mu_\nu$, where in this 
context, $\d$ is the Kronecker delta function.

Of all of the general coordinate systems that could be used to
characterize our flat Minkowski spacetime manifold, $M$, obviously the global 
$x^\a$ inertial frame coordinates considered in Sec.~\ref{basic} are a natural 
choice.  However, for a given path ensemble, other natural choices
also arise, based on quantities that we have already introduced. Let us
hereby define a system of {\em natural coordinates} as the 
curvilinear choice, 
\eb
	X^\mu = (c\,\l, {\bf C}) = (c\,\l, C^i) = (c\, \l, C^1, C^2, C^3), 
	\label{natcoord}
\ee
where $\l$ and $C^i$ are defined in Sec.~\ref{basic}. In that
section, these quantities were considered parameters; however, 
it is clear from the discussion therein that they can be promoted
to full-fledged coordinates, forming a good coordinate system
under the conditions discussed.  

\begin{figure*}
\includegraphics[scale=0.65]{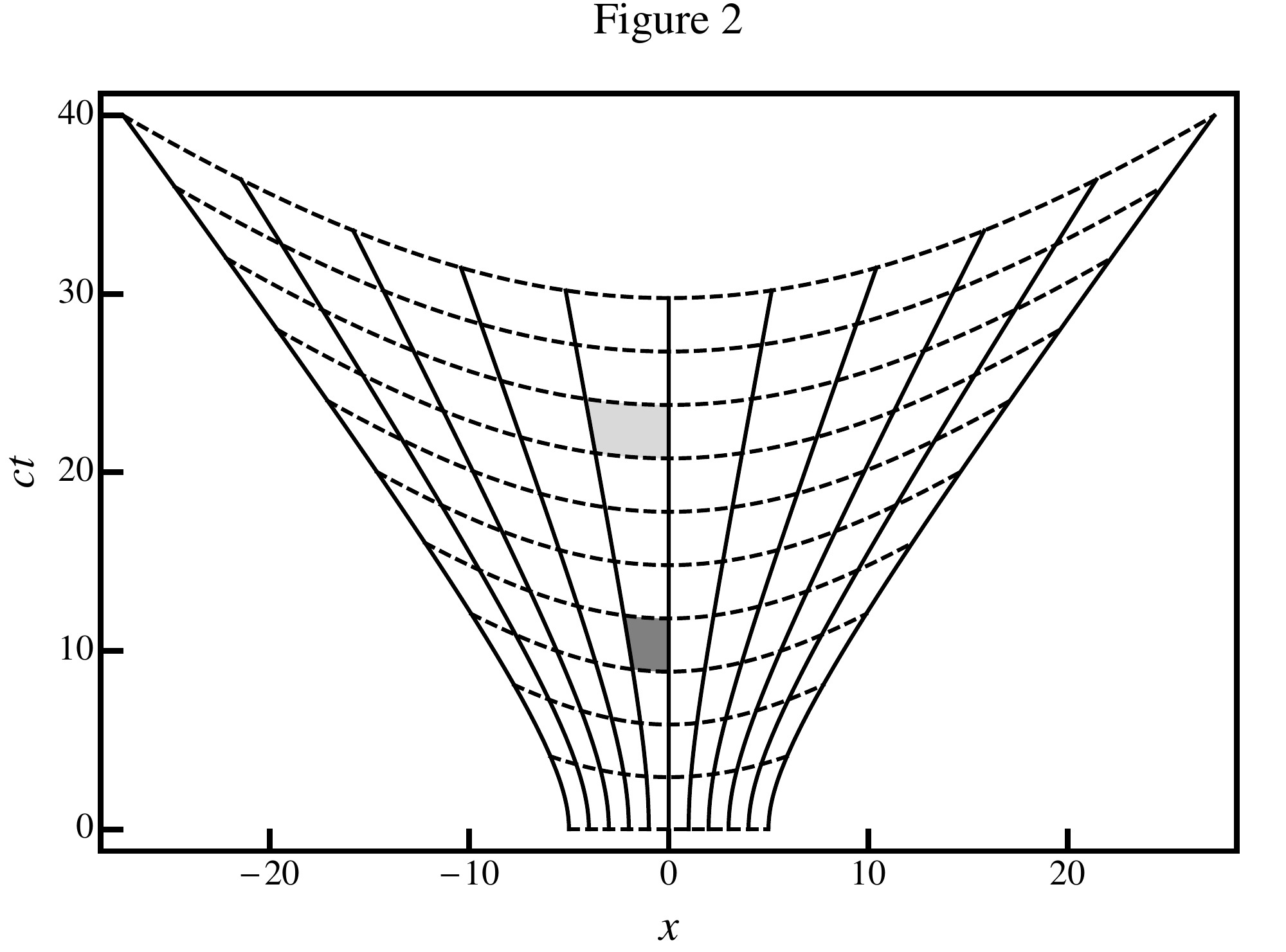}
        \caption{Quantum trajectories (solid curves) and simultaneity submanifolds (dashed curves)
for the 1d  $c\!\!=\!\!3$ relativistic Gaussian wavepacket of Sec.~\ref{gaussian}, as 
represented in the inertial coordinate frame, $x^\a=(c\,t,x)$.
At $t\!\!=\!\!0$ (i.e., along the $x$ axis), all spacetime events occur simultaneously. The
curvature of the simultaneity submanifolds at later times is an indication of
relativistic dynamical effects, which are quite pronounced.  The trajectories are everywhere
normal (orthogonal) to the simultaneity submanifolds. The dark gray and light gray 
patches correspond to those in Fig.~\ref{fig-natural}.}
        \label{fig-Gaussian}
\end{figure*}

\begin{figure}
\includegraphics[scale=0.65]{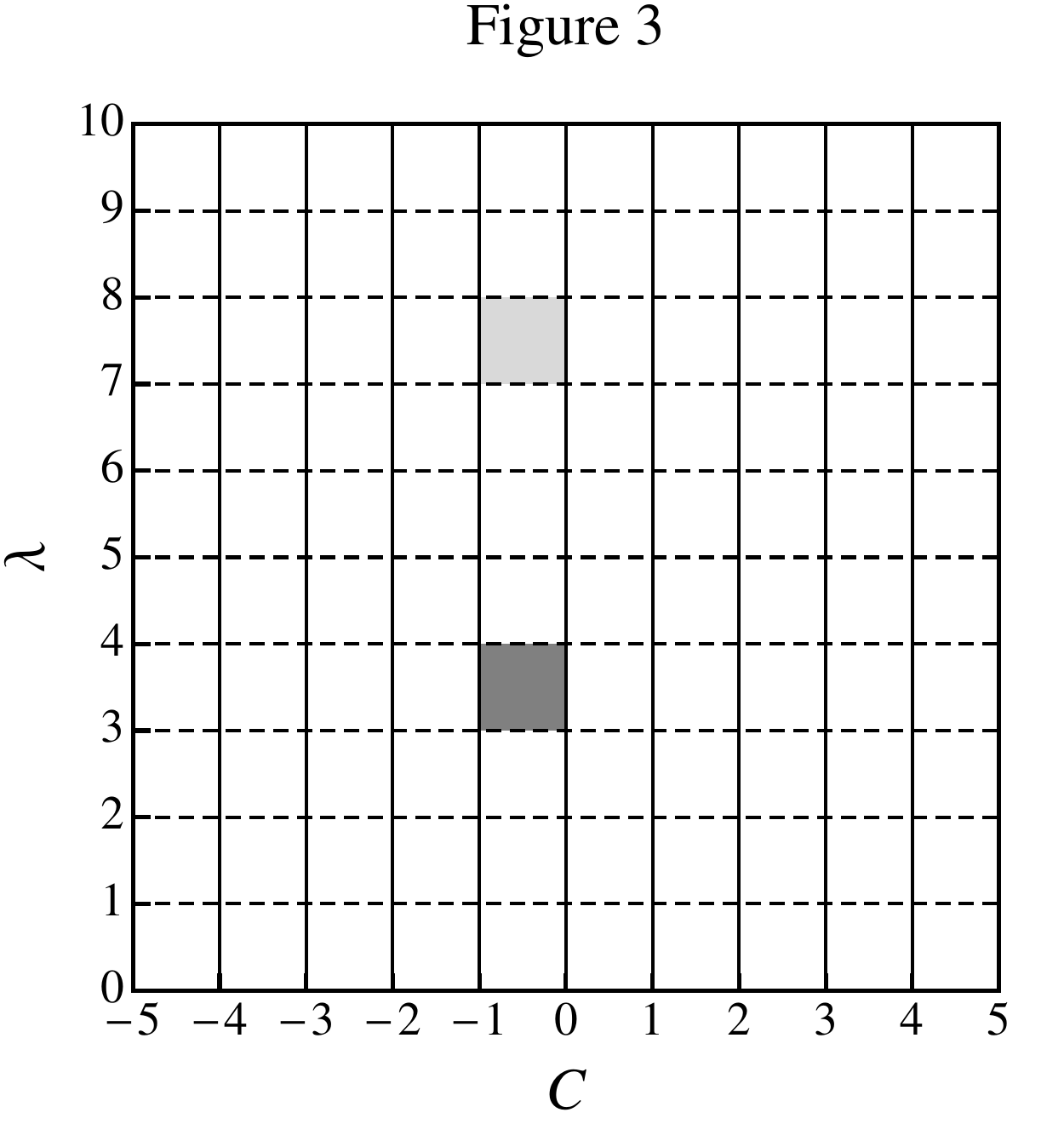}
        \caption{Quantum trajectories (solid curves) and simultaneity submanifolds (dashed curves)
for the 1d $c\!\!=\!\!3$ relativistic Gaussian wavepacket of Sec.~\ref{gaussian}, as 
represented in a natural coordinate frame, $X^\mu=(c \, \l , C)$. All
velocity and flux vectors point vertically (e.g., $U^\mu$), and all force 
vectors point horizontally (e.g., $f_Q^\mu$). The probability-length contained
in each gray patch is identical to that of the corresponding patch in Fig.~\ref{fig-Gaussian}, and
independent of the choice of $\l$ coordinate. This value is not the same for the
light gray and dark gray patches, although these two patches do have the same 
value for the spatial scalar probability density, $f(C)$.}
        \label{fig-natural}
\end{figure}

The physical meaning of the natural coordinates is such that
the intersection of the contours of the three functions, $C^i(x^\a)$, define
the individual paths within the ensemble, whereas $\lambda$ itself
describes the (ensemble) time evolution along a given path 
(see Figs.~\ref{fig-Gaussian} and~\ref{fig-natural}). 
Note that the introduction of a factor of $c$ into the definition of $X^0$
is consistent with the interpretation of $\l$ as a timelike coordinate.
Likewise, the $C^i$ coordinates, which (as we have seen) can be used
to label points on a given simultaneity submanifold, can be 
regarded as spacelike coordinates.  That is not to say, however, that $\l$ must
have units of time, and $C^i$ units of length.  
Rather, it is better to think of natural coordinates as being any choice
that {\em respects the natural division of $M$ into space and time} that is 
induced by a given path ensemble (as described in Sec.~\ref{basic}).  

Further justification for this interpretation of the natural coordinates
is provided by the fact that  for any coordinate transformation of a form
that respects this time/space division---i.e., the reparametrizations
$\l \ra {\l}' =  {\l}'(\l)$ and  ${\bf C} \ra {\bf C}' =  {\bf C}'({\bf C})$---the
new coordinates, $X^{\mu'} = (c\, \lambda', {\bf C}')$, are also seen
to be natural coordinates, by the definition given above. 
For the moment, we treat all such choices equally. Later, however,
after we have introduced suitable probabilistic structures and
dynamical elements on $M$,  we will find that a particular choice 
naturally emerges [i.e., $X^\mu = (c \,\T, {\bf P})$]. 

An important feature of any set of natural coordinates---consistent with the
above interpretation---is that the metric tensor $\tilde g$  be 
block-diagonal.  Thus, $g_{0i} = g_{i0} = \eta_{\a\b} {J ^{\a}}_{\!0}{J ^{\b}}_{\!i}=0$,
and so 
\eb
	\tilde g = \left( \matrix{ g_{00} & {\bf  0} \cr
	                                      {\bf 0}  & \tilde \g } \right).
	 \label{gblock}                                     
\ee  
In \eq{gblock} above, $\tilde\g$---the so-called ``spatial 
metric''---represents the $3\times3$ spatial 
block of the full metric tensor $\tilde g$. 
Natural coordinates are therefore orthogonal  with respect to 
the division of space and time (although the spacelike coordinates $C^i$ 
are not necessarily orthogonal amongst themselves). 
This is a manifestation of the fact that the timelike and spacelike subspaces 
of the tangent vector spaces were constructed, by design,
to be orthogonal to each other [\eq{orthosub}].  Note that reparametrizations
of $\l$ do not affect $\tilde\g$, and reparametrizations of ${\bf C}$ do not
affect $g_{00}$. 

The block-diagonality assertion above is readily proven in terms of the {\em inverse} 
metric tensor,
\ea{
	g^{\mu\nu}   & = &  \eta^{\a\b} {K ^{\mu}}_{\!\a}{K ^{\nu}}_{\!\b} \nonumber \\
	{\tilde g}^{-1} & = & {\tilde K} \cdot \tilde \eta \cdot {\tilde K}^T. \label{ginvdef}
}
Note that $ {K ^{0}}_{\!\a} \propto \p_\a \l$, 
whereas the contravariant vectors $ \eta^{\a\b} {K ^{i}}_{\!\b}$ describe displacements
within the simultaneity submanifolds.  Thus, from \eqs{orthosub}{ginvdef}, we have 
$g^{0i}=g^{i0}=0$, and the same must therefore be true for $\tilde g$ itself.  
Another useful set of relations to follow from the time/space
orthogonality of the natural coordinates is
\eb
     g^{00} = {1 \over g_{00}} = c^2  \eta^{\a\b} (\p_\a \l)(\p_\b \l) = - \of{{d\l \over d\t}}^2.
     \label{g00}
\ee
 The last equality relates the time-time component of the (inverse) metric tensor
 to $d\l / d \t$, the local relation between 
 ensemble and proper time (along a given trajectory/path).  
 
 Finally, we have a relation for the determinant of $\tilde g$, which will appear
 in various expressions. Following standard convention, we take 
 \eb
       g = \det \tilde g = - (\det \tilde J)^2 = -J^2= -1/K^2
       \label{gdet}
 \ee
 No absolute value is implied in \eq{gdet} above; indeed, we see that
 $g$ must be negative.  Equation~(\ref{gdet}) above is completely general.
 For natural coordinates, however, this reduces to 
 \eb
 	g = g_{00} \g, \qquad \rm{where} \quad \g = \det \tilde \g
 \ee
 (and $\g$ should not be confused with the index). 
 It is clear from the previous definitions that $g_{00} <0$
 and $\g >0$; thus, the -+++ signature of the inertial frame metric
 is retained, in keeping with our time/space interpretation of
 the natural coordinates.

\subsection{Probability in 3d}
\label{prob3d}

Classical statistical mechanics, trajectory-based quantum mechanics, and
relativistic hydrodynamics all include the notion of density fields that
propagate through time and 
space.\cite{holland,wyatt,poirier10nowave,poirier12nowaveJCP,poirier12ODE,carroll,weinberg}
The densities are associated with
corresponding flux vectors, that govern the local motion of fluid elements.
Densities and their corresponding fluxes obey a continuity relationship, 
expressing the physical conservation law for a given quantity---be it energy/mass, 
charge, probability, or number of particles. 

As we seek here to generalize the non-relativistic theory of quantum trajectory 
ensembles, and since the non-relativistic theory is mathematically equivalent to 
the TDSE, the conserved quantity in question can be taken to be probability. In the 
non-relativistic case, establishing the appropriate continuity equation is relatively 
straightforward---essentially because the time 
coordinate $t$ is uniquely determined. In the relativistic case, however, there are myriad 
probability-related quantities that can in principle be defined.  One can construct 
both scalar and vector densities (current or flux four-vectors), 
as well as stress-energy-type {\em tensors}---all of which can be further subcategorized 
as to whether they exist on  the full 4d spacetime, or only on submanifolds.  In addition, 
whereas density quantities are generally not true invariants (i.e., with respect to general 
coordinate transformations),\cite{carroll,weinberg} one can often construct invariant 
versions of these quantities.  To cut through the morass of possibilities, we apply the 
standard procedure of contemplating how such quantities, and  the relationships 
among them, should transform under various coordinate transformations.

We also rely on the following assumption, borrowed from the non-relativistic
theory:
\begin{itemize}
\item{{\em Postulate 1:} Probability is conserved along quantum trajectories.}
\end{itemize}
In the non-relativistic case, ``probability'' means the true, unitless,
probability element---i.e., the (spatial) probability density times the (spatial) volume 
element, $\rho(t,{\bf x}(t))\, d^3x(t)$.  Postulate 1 stipulates that along a given 
trajectory (i.e., for fixed $\bf C$),  this quantity is independent of $t$.  Under the 
coordinate transformation ${\bf x} \ra {\bf C}$ (for fixed $t$), the probability density 
transforms as 
\eb
	\rho(t, {\bf x}(t))\, d^3x = f(t,{\bf C})\, d^3C,
	\label{probtrans}
\ee
because the probability element itself must be a scalar invariant.  Since $\bf C$ and 
$d^3C$ are independent of $t$ along a trajectory, the probability conservation 
postulate therefore implies that $f(t,{\bf C}) = f({\bf C})$ is itself independent of $t$. 
Thus, $f({\bf C})\, d^3C$ assigns a definite probability element to each trajectory
in the ensemble, {\em for all time}.

We posit a similar situation in the relativistic case. Let $f({\bf C})$ be introduced 
as a probability density  on $\bf C$ space---i.e., on the set of individual  
paths within a given ensemble.  This density has units of $C^{-3}$, and is presumed 
to be normalized to unity:
\eb
	\int \! f({\bf C}) \, d^3 C = 1
	\label{probnorm}
\ee 
The value of $f({\bf C})$ must be independent of the value of $\l$, in order to satisfy the probability
conservation postulate---but equally importantly, it must be invariant of
$\l$ {\em reparametrizations}, $\l \ra \l'(\l)$.  The function $f({\bf C})$ can thus be 
``pulled back'' to the individual simultaneity submanifolds on $M$,\cite{carroll,weinberg} 
but in no sense should it be regarded as a probability density on $M$ itself.  We henceforth 
refer to $f({\bf C})$ as the ``spatial scalar probability density'' on ${\bf C}$. 

The situation is exemplified by Fig.~\ref{fig-natural}, which depicts the spacetime manifold $M$ as 
charted in a natural coordinate system. In this frame, individual paths appear as vertical 
lines, and simultaneity submanifolds (surfaces of constant $\l$) as horizontal lines.
The evolution of ensemble time---technically, the $\l$-parametrized family of 
$M \ra M$ diffeomorphisms generated by the $\l$ velocity vector field, 
\eb 
     V^\mu = {dX^\mu\over d\l} = (c,0,0,0)
     \label{Vdef}
\ee     
---serves to advance the simultaneity submanifolds
vertically through the spacetime $M$. Of course, the ``rate'' at which this occurs depends on the 
choice of ensemble time coordinate $\l$---which in turn affects the vertical length scale
in Fig.~\ref{fig-natural}, and the ``thickness'',  $c\, d\l$, of a four-volume element, $c\, d\l \,d^3C$.
The changes in these quantities brought about by a reparametrization
of $\l$ would in turn affect the values of any 4d density quantities on $M$.  
However, $\l$ reparametrization has
no effect  on the simultaneity submanifolds themselves, nor  on 
the 3d quantities that live on these submanifolds, such as  the spatial
scalar probability density, $f({\bf C})$, and the spatial volume element, $d^3C$.  

On the other hand, we  of  course demand that  $f({\bf C})$ should transform as a 
probability density under reparametrization of the {\em spatial} coordinates, 
${\bf C} \ra {\bf C}'$:
\eb
	f'({\bf C}') d^3C' = f({\bf C}) d^3 C \quad  ; \quad
	f'({\bf C}')  = \sqrt{{\g' \over \g}} f({\bf C}) \label{fCtrans}
\ee
Technically speaking, as a function of ${\bf C}$,  $f({\bf C})$ is not a scalar
invariant field, but  a 3d {\em scalar density of weight\/} $W\!=\!-1$. 

\subsection{Probability and flux in 4d}
\label{prob4d}

As indicated,  the 3d (spatial)  $f({\bf C})$ density transforms in a well-prescribed way, 
under all coordinate transforms that preserve the natural coordinate structure.  Our
ultimate goal, however, is a fully covariant formalism, for which {\em any} choice
of coordinates might in principle be used.  We should therefore also consider 
the appropriate 4d scalar probability density (and the corresponding flux 
four-vector).  A natural way to achieve this is through the use of the 
{\em invariant} form of the spatial scalar probability density---which 
is a true scalar invariant (weight $W\!\!=\!\!0$). We denote such invariant 
forms of density quantities with an asterisk superscript. The  ``spatial scalar 
invariant probability density'' is thus found to be 
\eb
         f^*({\bf C}) = {f({\bf C}) \over \sqrt{\g}},
         \label{invdens}
\ee 
as is evident from \eq{fCtrans}.

The advantage of working with true scalar invariants is that it is quite
straightforward to pull back a scalar invariant function on $M$ to any  
submanifold of $M$, simply by restricting the domain of the function 
accordingly.  In the present case, we would like to define a 4d scalar 
invariant probability density on all $M$, denoted $\rho^*(X^\mu)$.  What 
we have is a 3d spatial scalar invariant probability density $f^*$ defined on
individual simultaneity submanifolds; but since the set of all such 
submanifolds foliate $M$, $f^*$ can be promoted to an actual scalar invariant
function on $M$, i.e. $f^*(X^\mu)$. It is therefore completely natural to define
\eb
        \rho^*(X^\mu) = f^*(X^\mu)
        \label{rhostar}
\ee   
in a generally covariant manner.   The desired 4d scalar probability
density (of weight $W\!=\!-1$) then becomes
\eb
	\rho(X^\mu) = \sqrt{|g|}\, \rho^*(X^\mu) =  \sqrt{{|g| \over \g}}\, f\sof{{\bf C}(X^\mu)}
	\label{rhogen}
\ee 

Equation~(\ref{rhogen}) above is the completely general covariant expression.
In this context, however, it must be understood that $g$ refers to the completely
general coordinates $X^\mu$, whereas $\g$ refers to the coordinates $\bf C$.
In the special case where $X^\mu= (c \,\l, {\bf C})$ is taken to be the 
natural coordinate frame itself, then \eq{rhogen} reduces to 
\eb
       \rho(X^\mu) = \sqrt{-g_{00}}\ f({\bf C}) = \of{{d\t \over d\l}} f({\bf C}).
       \label{rhonat}
\ee
In any event, $\rho(X^\mu)$ behaves as a true 4d scalar density should;
 the element $\rho(X^\mu)\, d^4X$ is a scalar invariant. This
element is {\em not} unitless, however; it has units of length, as the invariant
form $\rho^*$ has units of length$^{-3}$. 

From the scalar probability density $\rho(X^\mu)$, 
we wish to construct a {\em flux four-vector}, $j^\mu$, such that 
\eb
	j_\mu\, j^\mu = g_{\mu\nu}\, j^\nu j^\mu = -c^2 \rho^2.
	\label{jcond}
\ee
Intuition (e.g., from electromagnetic current) suggests that $j^\mu$ should transform
as a contravariant four-vector, which would make  $\rho$ a scalar invariant. In
SR, this is completely correct, in the sense that $\rho'(x^{\a'}) = \rho(x^\a)$ is
unaltered under Lorentz transformations, because the volume element
$d^4x' = d^4x$ is unchanged. Thus,  $j^\mu$ indeed transforms
as a four-vector under \eq{LT}. In the completely general context however, incorporating 
transformations to curvilinear coordinate frames, one finds that $\rho$ and $j^\mu$ 
should transform, respectively, as scalar and contravariant vector {\em densities}, 
of weight $W\!=\!-1$.  This distinction is quite critical from the perspective 
of the covariant continuity equation (to be defined shortly).  

From \eqs{fourvcond}{jcond}, and the requisite transformation properties, the desired 
flux quantity is evidently
\eb
	j^\mu = \rho(X^\mu)\, U^\mu = f({\bf C})\, V^\mu,
	\label{genflux}
\ee
where the first equality is generally true, and the second holds
[from \eqs{Vdef}{rhonat}] in a natural coordinate frame.

\subsection{Covariant continuity}
\label{cov-continuity}

Next, we derive a continuity equation in a natural coordinate frame. We expect
this to take the form of a vanishing four-divergence of $j^\mu$, i.e.
\eb
	\p_\mu\, j^\mu = 0
	\label{continuity}
\ee
From \eqs{Vdef}{genflux}, the flux in a natural coordinate frame is given by
$j^\mu =  (c\, f({\bf C}), 0, 0,0)$, for which \eq{continuity} is 
clearly satisfied, by virtue of the probability conservation postulate.

Is \eq{continuity} also true for general coordinates? Since derivatives of vector quantities
are involved, ordinarily, one would have to replace the partial derivatives
in \eq{continuity} with covariant derivatives, in order to obtain a generally 
covariant expression. However, for the special case of a contravariant 
vector density with weight $W\!=\!-1$ (or a covariant vector density with weight $W\!\!=\!\!+1$)
it turns out that the {\em covariant four-divergence is equal to the usual (partial derivative) 
four-divergence}. 
Thus, $\p_\mu\, j^\mu = \nabla_\mu \,j^\mu$ holds in all coordinate frames,
and therefore, so does \eq{continuity}. We take this to be the 
{\em covariant continuity equation}. 

The four-divergence property described above can be proven as follows. 
Let  $j^\mu$ be a  contravariant vector density of weight $W\!=\!-1$. 
The standard expression for the covariant derivative is
\eb
	\nabla_\nu\, j^\mu = \p_\nu\, j^\mu + \G^\mu_{\nu\k}\, j^\k  - \G^\k_{\nu\k} \,j^\mu,
	\label{codvecdens}
\ee 
where the $\G^\mu_{\nu\k} $ are the famous Christoffel symbols,\cite{carroll,weinberg}
defining the (metric compatible) connection:
\eb
	\G^\mu_{\nu\k} = {1 \over 2} g^{\mu\s} (\p_\nu g_{\k\s} + \p_\k g_{\s \nu} - \p_\s g_{\nu \k})
\ee
To obtain the covariant divergence from \eq{codvecdens}, $\nu$ is set equal to $\mu$, 
which then becomes a dummy index that is summed over (Einstein notation). By the symmetry
of the Christoffel symbols with respect to their lower indices, the second and third 
terms in \eq{codvecdens} then cancel, leaving the desired equality. 

The covariant continuity equation is easiest to interpret in a natural coordinate frame.
Here, the $j^\mu$ vector field is ``vertical'' (Fig.~\ref{fig-natural}), and simply describes how probability density 
is carried along under the action of the $V^\mu$ velocity field.  It is important to note that it is
the {\em spatial scalar probability density}, $f({\bf C})$, that is conserved in this sense,
not the scalar probability density $\rho(X^\mu)$ per se. To within a constant factor of $c$, 
the former is just the zeroth component of the flux four-vector, $j^0$, in a natural coordinate
frame.  More generally, i.e. in an arbitrary coordinate system, it is $(j^0/c)$ that we ordinarily 
think of as the ``probability density''.    In particular, if $X^\mu$ is any coordinate system 
for which $X^0$ and $X^i$ are orthogonal, then \eq{continuity} and the 
generalized Stokes theorem\cite{carroll} imply that
\eb
	\int j^0(X^\mu)\, dX^1 dX^2 dX^3 = {\rm const,}\qquad {\rm for\  all}\,\,X^0
	\label{Pcons}
\ee
[provided $j^0(X^\mu)$ vanishes asymptotically]. 

Note that the orthogonal condition above is {\em not} the same as the natural
coordinate condition, and is in fact far less restrictive.  For instance (and in addition to 
any natural coordinate frame) it includes any inertial coordinate frame, $X^\mu = x^\a$, 
in which case \eqs{continuity}{Pcons} reduce to the usual SR forms. 
The inertial frame forms of the density and flux quantities bear further discussion.
From \eq{rhogen}, we find that
\eb
	\rho(x^\a) = \rho^*(x^\a) =  {f\!\sof{{\bf C}(x^\a)} \over \sqrt{\g}}.
	\label{xdens}
\ee
Likewise, from \eqs{rhonat}{genflux}, we obtain
\ea{
	j^\a & = &  {f\!\sof{{\bf C}} \over \sqrt{\g}}\,U^\a = 
	c\, {f\!\sof{{\bf C}} \over \sqrt{-g_{00} \g}}\,{J^\a}_{\!0}    \nonumber \\
	\quad j^0 & = & c \, {f\!\sof{{\bf C}} \over \sqrt{\g}}\,  {dt \over d\t}
}
Clearly, $\rho(x^\a)$ and $(j^{\a}/c)$ both have units of length$^{-3}$, 
as they should.  Note that $g_{00}$, $\g$, and ${J^\a}_{\!0}$ refer to the
natural coordinate frame.

\subsection{ Uniformizing natural coordinates}
\label{uniformizing}

Let us return to a natural coordinate frame, and consider the ensemble time evolution, as 
indicated in Figs.~\ref{fig-Gaussian} and~\ref{fig-natural}. 
Although the probability element $\rho(X^\mu)\, d^4X$ is not conserved
along trajectories, it is invariant.  Thus, the probability-length contained within 
the dark gray patches in Fig.~\ref{fig-Gaussian} and Fig.~\ref{fig-natural} is the same, and the probability-length 
within the light gray patches  in Fig.~\ref{fig-Gaussian} and Fig.~\ref{fig-natural} is the same, but the value
for the dark gray patches is {\em not} the same as for the light gray patches. 
Fundamentally, this is because
$\l$ and $\t$ are generally incompatible, and so $d\t/d\l$ must vary across $M$.
Put another way, a reparametrization of $\l$ could expand the vertical scale 
of the light gray patches  without necessarily changing that of the dark gray patches,
which has a corresponding effect on the $\rho$ values. 
Clearly, $\rho$ cannot be conserved for a general choice of 
ensemble time, $\l$. 

Nevertheless, it would be nice if we could define a ``proper-time like'' 
reparametrization of $\l$, for which $\rho(X^\mu)$ was as close to being
conserved as possible.   The best we can manage along these lines
is to choose a  $\l$ such that the {\em spatial integral}
of $\rho(X^\mu)$, i.e.
\eb
	\int \!\rho(X^\mu) \, d^3C = \int {d\t \over d\l} \, f({\bf C})\, d^3C
\ee
is conserved over time.   We can therefore define, as an ensemble proper time,
$\T$, that choice of ensemble time $\l$ for which 
\eb
        \int\! {d\t \over d\T} \, f({\bf C})\, d^3C = 1, \qquad {\rm for\  all}\,\,\T.
        \label{Tdef}
\ee  
  
The interpretation of \eq{Tdef} is straightforward. By multiplying both sides
by $d\T$, we see that,  at any time $\T$, the interval $d\T$  is just the {\em path ensemble
average} of the proper time interval $d\t$.  If the $d\t$ intervals happen to be the same
for all paths in the ensemble at a given $\T$, then $d\T=d\t$ at that time.  If this is 
true at all times, then $\T=\t$---as describes  the special case of quantum inertial observers.  
In general though, the $d\t$ values differ across the ensemble, and so 
$d\T$ is obtained as the probability-weighted average of the $d\t$ values for the 
individual paths.  Note that for a given path ensemble, the $\T$ coordinate as 
specified here is  {\em uniquely defined}, to within an overall additive constant.

As reasonable as the above prescription may appear, \eq{Tdef} is not the 
most natural choice for an ensemble proper time, $\T$; that choice will be 
introduced in Sec.~\ref{conversion}.  For one thing, even with \eq{Tdef}, the probability-length
of the dark gray patches in Figs.~\ref{fig-Gaussian} and~\ref{fig-natural}
 is still not equal to that of the light gray patches---although the spatial integrals across 
 the simultaneity submanifolds are now equal.
The ensemble proper time as defined above can be regarded as a ``uniformizing 
coordinate,'' in the sense that it does the best possible job of  spreading out the scalar
probability density, $\rho(X^\mu)$, uniformly throughout (ensemble) time.  

One might also consider a similar reparametrization of the (natural) spatial coordinates, 
${\bf C}$. Here, it is possible to perfectly uniformize the probability 
distribution $f({\bf C})$, and in fact, exactly this procedure has been used in the 
previous non-relativistic work.\cite{poirier12nowaveJCP}  
We hereby denote such a uniformizing choice of spatial coordinates for $\bf C$ 
as $\bf C\!=\!P$,  defined such that
\eb
	f({\bf P}) = 1.\label{Pdef}
\ee	
For a single spatial dimension, the $P$ coordinate is uniquely determined,
apart from an additive constant (which fixes the range of allowed $P$ values);
it can be interpreted as the total (integrated) probability that exists to the left
of a given path in the ensemble (hence the nomenclature ``$P$,'' often used
in such contexts).  In higher dimensional spaces, the $\bf P$ coordinates are
also uniquely  determined---apart from (spatial) volume-preserving diffeomorphisms.  

With $\T$ taken to be any ensemble proper time [e.g., either that of \eq{Tdef} 
or of Sec.~\ref{conversion}],  we regard  $X^\mu = (c\,\T, {\bf P})$
to comprise a set of ``uniformizing natural coordinates,'' in terms of which 
$\T$ has units of time, ${\bf P}$ is unitless, $f({\bf P}) = 1$ is unitless, and
$\rho(X^\mu) = (d\t / d\T) f/\sqrt{\g}$ has units of length$^{-3}$ (as for the 
inertial frame case).


\section{Dynamical  Considerations}
\label{dynamical}

We now have many of the elements in place that we need to develop
a trajectory-based theory of relativistic quantum dynamics.  However,
there are still a few more issues, dynamical in nature, that must first
be addressed. In the first two subsections below, we consider 
single-trajectory classical dynamics, both
non-relativistic and relativistic.  Even in this context, 
there are certain subtle aspects that turn out to have extremely
important ramifications for the relativistic quantum case.  In the third
subsection below, we consider the trajectory-based non-relativistic quantum 
theory, presenting some key results from earlier work, and setting the stage
for addressing quantum effects in the relativistic context.

\subsection{Classical dynamical considerations}
\label{classical}

Consider a single, non-relativistic classical particle. The dynamics
are described by the trajectory, $x^l(t)$.  The velocity vector is ${\dot x}^l=dx^l/dt$,
and the non-relativistic classical equation of motion is
\eb
	m {\ddot x}^l = - \prt {V(t,{\bf x})}{x^l} = f^l,
	\label{Newton} 
\ee
where $V(t, {\bf x})$ is the potential energy field, and $f^l$ is the force vector. 
Note that we are considering the general, nonconservative case where $V$ 
may depend on $t$ as well as ${\bf x}$.  The reason is that we wish to 
adopt a 4d spacetime viewpoint, even in this non-relativistic context.
We therefore continue to use notation such as ``$x^\a$'' and ``$g_{\a\b}$,''
mostly without ambiguity. Thus, for example, the Euclidean metric, 
 $g_{\a\b} = \delta_{\a\b}$ is presumed. 

From this 4d vantage, a striking feature of \eq{Newton} is that  the force vector
{\em does not include a time component}, even when $V$ depends on $t$. 
In other words, the dynamical equations do not make use of what would be
the entire force four-vector, $f^\a=f_\a = \p_\a V$, but only the spatial components 
of this vector---i.e., the projection onto the (non-relativistic) simultaneity 
submanifold (just 3d space itself, ${\bf x}$).  If the
time component {\em were} used, it would lead to an additional dynamical
equation of the form
\eb
	m {\ddot t} = f^0 \propto - \prt{V}{t},
\ee
which is manifestly incorrect for time-dependent potentials, because
$ {\ddot t} = d{\dot t} / dt = d (1)/dt = 0$.

From a physical standpoint, the restriction to just the spatial components
of the force vector is not surprising. Suppose that the partial time derivative of 
$V(t,{\bf x})$ {\em were} employed 
in the determination of the dynamical force.  That would mean that this force, 
which drives the future time evolution of the particle, would itself depend
on the {\em future} states of that particle---a highly untenable situation 
vis-\`a-vis causality, which can be reasonably dismissed.  Of course, time reversibility
implies that the force must also be independent of the 
{\em past} state of the particle, which leaves only the present.
In this manner, we are led to:
\begin{itemize}
\item{{\em Postulate 2:} All force vectors, together with the quantities
used in their construction,  must ``live'' on the simultaneity submanifolds.}
\end{itemize}
The meaning of this statement will be made more precise as we go along. In any case,
we take Postulate 2 as a necessary condition for any viable physical theory---at least for 
particles with mass, in the context of non-relativistic or relativistic classical or 
quantum mechanics.  As shown above, it certainly holds for non-relativistic classical particles.  

It also holds for {\em relativistic} classical particles, as we now demonstrate. 
Of course, it is not possible to define a global simultaneity submanifold
from a single relativistic trajectory, $x^\a (\t)$.  All that is required for the present 
purpose, though, is a {\em local} specification of simultaneity, which we do in fact 
have.  As discussed in Sec.~\ref{basic},  this is found---for a given point $p$---in the orthogonal
subspace of $T_p$ that is orthogonal to $U^\a$ at $p$.  
According to Postulate~2, the relativistic force vector $f^\a$ must belong to this 
orthogonal subspace. That this is the case can be shown 
using the relativistic classical equation of motion:
\eb
	m {d^2 x^\a  \over d \t^2} =  m {d U^\a \over d\t} =  f^\a
	\label{reldyn}
\ee  
(note that we are still working in an inertial frame). 
From \eqs{fourvcond}{reldyn},
\eb
	{m \over 2} \,{d (\eta_{\a\b}\, U^\a U^\b) \over d \t} =  \eta_{\a\b}\, U^\a f^\b = 0. 
	\label{relfcond}
\ee
Thus, indeed, $f^\a$ belongs to the orthogonal subspace, and 
hence points along the simultaneity submanifold.

Let us pause to consider the ramifications of  \eq{relfcond}. Since the orthogonal
subspace at a given point $p$ depends on the 
trajectory, this implies that $f^\b$ itself must also depend on the trajectory
at $p$---i.e., not just on the point itself. Thus, for example, a relativistic force vector 
can  {\em never} be simply the gradient of some scalar invariant potential: 
$f^\a \ne - \eta^{\a\b} \p_\b \Phi$.   On the other hand, one simple way that 
\eq{relfcond} can be achieved is by taking $f^\a$ to be the product of an 
antisymmetric tensor and $U^\a$.  This is the general form for the relativistic
force that ensues when the Lagrangian includes a term that is the scalar product of
$U^\a$ and some four-vector field.  The classic example is 
the vector potential, $A_\a$, of electromagnetic theory, for which the resultant 
antisymmetric tensor, $F_{\a\b}$, is the electromagnetic field strength tensor.  
Thus, the  ``velocity-dependent''  nature of a relativistic force 
field---well-known in the electromagnetic context---is seen to be a general 
requirement arising from Postulate~2.

Note that,  whereas $f^\a$ does indeed belong to the orthogonal subspace of
$T_p$, the same cannot be said for $A_\a$ and $U^\a$---despite the fact that 
 these  are clearly ``elements used in the construction'' of $f^\a$.  
 Postulate 2 can still be regarded to be satisfied---albeit in a weaker sense
 than for $f^\a$ itself---because  the $A_\a$ and $U^\a$ vectors belong to the 
 tangent space $T_p$ for the point $p$ itself, and not some other point.  Thus, 
 in that sense, these vector quantities belong to the simultaneity submanifold.  
The main point,  though, is that  $f^\a$ should in no sense depend on quantities
associated with the {\em future} (or past) particle states.  Taking $p$ and $U^\a$ 
together as constituting the local state of the particle,  then the determination of 
$f^\a$ at $p$ should not depend on future or past values for these quantities, 
neither explicitly nor implicitly. 

Technically speaking, it is the $F_{\a\b}$ tensor, rather than $A_\a$ itself, 
that is used in the construction of $f^\a$.  As it happens, the $F_{\a\b}$ tensor 
{\em does} depend on future particle states, because this quantity is obtained 
from all partial derivatives of $A_\a$.  Thus, it is not only $A_\a$ at
the point $p$ itself that is consulted,  but also, the value 
of $A_\a$ at nearby points, displaced from $p$ in all four spacetime directions.  
This situation might appear to violate Postulate 2, but in fact, it does not. 
The reason is that the full $F_{\a\b}$  tensor is not actually required to construct 
$f^\a$---only that portion of the tensor that is projected onto the  local 
simultaneity submanifold is used. Consequently, although $A_\a$ values at points
other than $p$ are  required, these points  occur simultaneously with $p$, 
lying neither in its past nor its future,  and so Postulate 2 is indeed satisfied.  
Similar arguments will prove relevant, also, for  quantum force fields,  
although these arise in an altogether different manner. 

A more precise statement of Postulate 2 also requires a clearer specification
of the dynamical ``evolution coordinate,''  denoted $\xi$. 
Under arbitrary reparametrizations, $\xi \ra \xi'(\xi)$, the forms of \eqs{Newton}{reldyn} 
are not preserved, and---depending on interpretation---``time'' components of the 
resultant force vectors do in fact arise. However, these should be properly 
regarded as fictitious.   The fact is that the
metric tensor, together with the local simultaneity submanifolds, give
rise to a natural local choice for $\xi$, in terms of which Postulate 2 may be expected
to hold.  This choice is such that $\xi=\rm{const}$ within the simultaneity submanifold,
and $d\xi / d s = \rm{const}$  for displacements 
perpendicular to the simultaneity submanifold (where $ds^2$ 
is the line element, defined in Sec.~\ref{preliminaries}).  This specification of the evolution
coordinate $\xi$ may be applied to both the non-relativistic and relativistic 
cases considered above---giving rise to  $\xi = t$ and $\xi = \tau$, respectively
(more generally, $\xi$ may also include affine transformations of the above forms).  
With these choices for $\xi$, we have already seen that Postulate 2 holds for both 
non-relativistic and relativistic classical mechanics.

\subsection{Arbitrary time reparametrizations, as applied to classical electromagnetism}
\label{electromagnetism}

As the canonical example of a relativistic classical force field with the above 
velocity-dependent form, we consider the case of electromagnetism in
some detail.  This will turn out to be of particular benefit for the development
of relativistic quantum force fields, in at least two ways.  First, it provides
an understanding of how potential energy contributions, in general, should 
enter into the relativistic Lagrangian.  The second benefit emerges from 
a reworking of  the electromagnetic example in terms of an 
arbitrary time reparametrization, i.e. $\t \ra \l(\t)$, which is not usually seen 
in standard treatments.  In the electromagnetic
context, this introduces unnecessary complexity, but it provides
important insight into how to handle the relativistic quantum case---for which, 
as we have seen in Sec.~\ref{general-prob}, ensemble time $\l$ and proper time $\t$ are 
incompatible. 

Let us start with the simplest case of a free particle. Let $x^\a (\l)$ denote
an arbitrary timelike path, described by the arbitrary parameter $\l$.  For 
fixed endpoints, the solution trajectory is that which maximizes the 
elapsed proper time, $\Delta \t$---i.e., it is a geodesic.  From \eq{dtaudef},
the elapsed proper time along any  path is given by 
\eb
      \Delta \t  =  \int  \!\of{{d\t \over d\l}}d\l =   \int \!\sqrt{- {1 \over c^2}\, \eta_{\a\b}\, {dx^\a \over d \l} \, 
                  {dx^\b \over d\l}} \, d\l.
       \label{freelam}           
\ee
One standard choice for $\l$  is the proper time, $\t$, itself.  
Making this choice, and multiplying \eq{freelam} by the negative rest 
energy, $-mc^2$, the maximization of $\Delta \t$  is converted into  the
equivalent minimization of the action,
\eb
	S = \int\! L\!\sof{x^\a_\t} d\t =  \int (-mc^2) \sqrt{-  {1 \over c^2}\, \eta_{\a\b}\, x^\a_\t \, 
                  x^\b_\t} \, d\t ,
                  \label{freetau}
\ee
where $x^\a_\t$ denotes $d x^\a / d \t  = U^\a$, and $L[x^\a_\t]$ is the
Lagrangian for the parameter choice, $\l=\t$.

Regardless of the particular choice of $\l$, the form on the right hand side of 
\eq{freelam} should be used in the Lagrangian. This is because it exhibits explicit 
dependences on $x^\a_\l = dx^\a/d\l = V^\a$, and thus gives rise to an 
Euler-Lagrange ordinary differential equation (ODE) for the solution trajectory, 
$x^\a(\l)$, in terms of the first and second $\l$ derivatives. Specifically, 
action minimization leads to the following Euler-Lagrange form:
\eb
      \prt L {x^\a}	- {d \over d\l} \of{{\prt L {x^\a_\l}}} = 0 \label{ELODE}
\ee
For the standard choice $\l=\t$, and the flat Minkowski spacetime metric
presumed here, the result is \eq{reldyn} (with $f^\a=0$).  For curved spacetime
manifolds, or flat manifolds described via curvilinear coordinates, 
$X^\mu$, taking $\l=\t$ leads to the standard geodesic 
equation,\cite{carroll,weinberg}
\eb
        X^\mu_{\t\t} + \G^\mu_{\nu\k}\, X^\nu_\t \, X^\k_\t = 0.
        \label{geo}
\ee

To a large extent, the above procedure is straightforward because of the 
presumed $\l = \t$ form, and the constraint that that induces. Even in this context, 
there are some ambiguities that can arise. Specifically, the square
root quantity in \eq{freetau} is necessarily equal to one.
Why not simply replace this expression with $1$, or otherwise multiply or 
divide $L[x^\a_\t]$ by as many such  square root factors as we wish?  
Likewise, incorporating the $\l=\t$ constraint explicitly into the optimization  
using the theory of Lagrange multipliers,  one should be able to add or subtract
additional terms of this form.  Even for the free particle 
case, this can lead to trouble---e.g., to $S= \int (-mc^2) \,d \t$, which is useless
from the point of view of generating {\em any} ODE for $x^\a(\t)$, much less the correct one. 
The situation is even more delicate, however, when a potential energy contribution 
is introduced into the Lagrangian. 

Consider, then, the standard form of the Lagrangian (in $\t$) for a single 
relativistic particle with charge $q$, acted on by the electromagnetic 
vector potential, $A_\a$:
\eb
     L\!\sof{x^\a, x^\a_\t} =   (-mc^2) \sqrt{-  {1 \over c^2}\, \eta_{\a\b}\, dx^\a_\t \, 
                  dx^\b_\t} + {q \over c} \, x^\a_\t \, A_\a  
                  \label{EMtau}
\ee
The potential energy contribution [i.e., the second term on the
right hand side of \eq{EMtau}] introduces an explicit Lagrangian dependence
on the coordinates $x^\a$, through the $A_\a$ vector field. The
Lagrangian form of \eq{EMtau} leads via \eq{ELODE} to the correct 
electromagnetic ODE---i.e., \eq{reldyn}, with 
\eb
	f^\a = {q \over c}\, \eta^{\a \b} F_{\b \g}\, x^\g_\t \quad {\rm and} \quad
	F_{\b \g}  = \p_\b A_\g - \p_\g A_\b.
	\label{EMdyn}
\ee

The above derivation requires explicit substitution of the \eq{dtaudef} 
equality---but only {\em after} the Lagrangian partial derivatives of
 \eq{ELODE} have been applied. Note that the one factor of the square root 
 quantity in the first term of \eq{EMtau} is necessary to generate the 
 kinetic energy contribution in the resultant equation of motion. 
 The potential energy term  of the Lagrangian does {\em not} include 
 this square root factor; if it did, the resultant ODE 
would be incorrect.  Likewise, any additional factors of the square root quantity 
in the first (kinetic) term of $L$ would also lead to incorrect ODEs. 

We will apply these findings to relativistic quantum force fields, but first we must 
also consider the general $\l$ case explicitly.  This can be regarded in terms of 
the coordinate transformation, $\t \ra \l(\t)$.  Since the action, $S$, must be a scalar 
invariant, $L$ must transform as a scalar density of weight $W\!=\!-1$---leading 
at once to the following form for the Lagrangian in $\l$:
\eb
     L\!\sof{x^\a, x^\a_\l} =   (-mc^2) \sqrt{-  {1 \over c^2}\, \eta_{\a\b}\, x^\a_\l \, 
                  x^\b_\l} + {q \over c} \, x^\a_\l \, A_\a  
                  \label{EMlam}
\ee

At this point, we must distinguish between two possibilities:  the case
where the explicit relation $\l(\t)$ is known {\em a priori}, vs. the case where it
is not.   In the former case, $\l(\t)$ implies an explicit 
constraint, analogous to (but more complicated than) that of \eq{dtaudef} for $\l=\t$.  
One can therefore apply a procedure similar to that used for $\l=\t$---i.e., the 
explicit constraint is invoked only after the Lagrangian partial derivatives 
have been evaluated.  The resultant ODE is the general-$\l$ version of 
\eq{reldyn}, i.e., 
\eb  
m\, x^\a_{\l\l} =   \t_\l^2\, f^\a + m\, {\t_{\l\l} \over \t_\l}\, x^\a_\l,
  	\label{reldynlam}
\ee 
where $\t_\l = d\t / d\l$, and $f^\a$ is from \eq{EMdyn}.  
That this form is correct may be readily verified---e.g.,  for the choice 
$\l = t = x^0/c$, in terms of which the classical electromagnetic 
equations of motion are very well known.

Let us now imagine---as is the case in the quantum context---that
the relation $\l(\t)$ [or the inverse, $\t(\l)$] is {\em not} necessarily known 
{\em a priori}.   This situation is evidently problematic, from the perspective
of the $\l$ evolution of \eq{reldynlam}.  At the initial time, it is
straightforward enough to specify initial values for $\t_\l$ and $\t_{\l\l}$---or
equivalently, $\l_\t$ and $\l_{\t\t}$.  However, there are many functions
$\l(\t)$ that share the same initial conditions, any one of which might in principle
be legitimately used; so how does the ODE ``know'' which choice to make, 
during the course of the propagation, if this is not already specified {\em a priori}? 

To address this question, the only recourse is to use the generic
relation for $\t_\l$, as implied by \eq{freelam}.  Substituting this form into
\eq{reldynlam}---or equivalently, deriving the Euler-Lagrange ODE
from \eq{EMlam} without applying any explicit constraint---one obtains
\eb  
m\, x^\a_{\l\l} =  \of{-  {1 \over c^2}\, \eta_{\b\g}\, x^\b_\l \, 
                  x^\g_\l } f^\a, 
  	\label{reldynlamgen}
\ee 
which is equivalent to \eq{reldynlam}, but without the last term. Thus,
$\t_{\l\l}= \l_{\t\t}=0$, or $\l(\t) = A \t + B$.  In other words, this 
procedure automatically picks out an {\em affine} relation for $\l(\t)$,
if no relation is specified {\em a priori}. (The values of the constants $A$ and $B$ 
are uniquely determined from the initial values for $\l$ and $\l_\t$, which are
arbitrary). 

All of the above is perfectly consistent---albeit unnecessarily complicated---in 
the context of relativistic electrodynamics.  The situation does not bode so well, 
however, for the relativistic quantum dynamical case.  Here, we know that 
$\l$  and $\t$ do not satisfy any global functional relationship of the form 
$\l = \l(\t)$---much less an affine relation.  Locally---and along a single given 
trajectory of the ensemble---such a relation can of course be established, 
although  it is not necessarily known  {\em a priori}, even in this context.  
All of this suggests that, in the quantum case, we must proceed with caution.

\subsection{Non-relativistic  quantum dynamical considerations}
\label{nonrel}

The trajectory-based formulation for  non-relativistic quantum dynamics
is, by this point, fairly well established.\cite{bouda03,holland05,poirier10nowave,holland10,poirier11nowaveCCP6,poirier12nowaveJCP,poirier12ODE}
  Here, we review only those
features that are most relevant for the relativistic generalization of Sec.~\ref{relativistic}. 
 Also, whereas our previous work has expressed the requisite quantities in terms of the
Jacobian matrices, ${J^{l}}_{\!i}$ and ${K^{i}}_{\!l}$, here, we adopt a tensorial,
generally covariant viewpoint, relying on the (spatial) metric tensor 
$\g_{ij}$, and its determinant, $\g$.

The non-relativistic formulation makes use of a spatial coordinate transformation 
from $x^l$ to $C^i$, for which the time coordinate, $t$, is unaffected. As in
the relativistic case, the $C^i$ serve as trajectory labeling coordinates for
a trajectory (or path) ensemble.  As discussed in Sec.~\ref{classical}, the simultaneity 
submanifolds are defined by surfaces of constant $t$ in the 4d non-relativistic 
spacetime manifold, $M$.  The $x^l \ra C^i$ transformation is thus parametrized by $t$,
with a complete solution trajectory ensemble taking the form $x^l(t, {\bf C})$.
In the $x^l$ coordinate frame, the metric tensor is presumed to be the usual 
Euclidean one, $\d_{lm}$, whereas in the $C^i$ coordinates, the metric tensor 
is denoted $\g_{ij}$. The latter is a function of both space $({\bf C})$  and time ($t$). 

As in the relativistic case, Postulate 1 is presumed (or, it can be
regarded as a logical consequence of Bohmian 
mechanics\cite{madelung26,bohm52a,bohm52b,holland,wyatt,durr92,berndl95}). 
The scalar probability density in ${\bf C}$ space, $f({\bf C})$, is thus independent
of $t$, and \eq{probnorm} is presumed to hold. From \eq{probtrans}, we have
\eb
	R(t, {\bf x}) = {f({\bf C})^{1/2} \over \g^{1/4}},
	\label{Rdef}
\ee 
where $R(t, {\bf x}) = \rho(t, {\bf x})^{1/2}$ is the usual wavefunction
amplitude, $|\Psi|$---i.e., the square root of the usual spatial probability density,  
$\rho(t, {\bf x})= \Psi^*\Psi$. The latter is a 3d density akin to $j^0/c$ from 
Sec.~\ref{cov-continuity}, 
but in the non-relativistic context,  it can also be interpreted as the 4d scalar 
probability density of Sec.~\ref{prob4d}.

We expect a scalar invariant action, $S$, of  the form
\ea{
	S & = &  \int \! \L[x^l,{\dot x}^l, \p_i x^l,\ldots]  \, d^3C\, dt  \nonumber \\
	    & =  &
	 \int \! L[x^l,{\dot x}^l, \p_i x^l,\ldots]  f({\bf C}) \, d^3C\, dt, 
	\label{nonrelaction}
}   
where ${\dot x}^l = (\p x^l /\p t)| {\bf C}$ and $\p_i = \p /\p C^i$.
The quantity  $\L$ is the Lagrangian density (a spatial scalar density of 
weight $W\!=\!-1$), whereas $L$ is a scalar invariant quantity (with units of
energy) that we refer to simply as ``the Lagrangian'' (although technically, 
that term should be applied to the $d^3C$ integral of $\L$.)  
The Lagrangian is a true scalar invariant (weight $W\!=\!0$). 
In non-relativistic quantum mechanics, it takes the form
\ea{
      L &  =  & {1\over 2} m\, \d_{lm}\, {\dot x}^l\, {\dot x}^m - V({\bf x})  
      -   \label{nonrelLagrange}  \\
      & & {\hbar^2 \over 8m} \sof{f({\bf C}) \over \g^{1/2}}^{-2} \!\!
       \g^{ij}\, \p_i\!\!\sof{f({\bf C}) \over \g^{1/2}} \p_j\!\!\sof{f({\bf C}) \over \g^{1/2}}.
      \nonumber
}
The metric tensor $\tilde \g$ depends explicitly on the $\p_i x^l$; the specific 
Lagrangian of \eq{nonrelLagrange} is therefore second order in the $C^i$
derivatives of $x^l$ (although an alternate, higher order choice of $L$ 
will also be considered).  

 The last term in \eq{nonrelLagrange} above is the (scalar invariant) quantum 
 contribution to the Lagrangian, denoted $-L_Q$.  This  term accounts for all 
 interaction or ``communication'' across trajectories---and hence, for all quantum 
 dynamical effects.\cite{holland05,holland10,poirier12nowaveJCP} 
 It can be expressed in a variety of ways, with the 
 \eq{nonrelLagrange} form above being particularly explicit.     
Note that Eqs.~(\ref{invdens}), (\ref{xdens}), and (\ref{Rdef}) imply that the quantity 
in square brackets can be interpreted as the spatial scalar invariant probability 
density, $f^*$---or equivalently, as the usual  $\rho(t,{\bf x}) = R(t, {\bf x})^2$.
Because $f^*$ is a scalar invariant, the partial derivatives $\p_i$
in \eq{nonrelLagrange} may be replaced with the corresponding covariant 
derivatives, $\P_i$.  Also,  since ($\P_i f) /f$ is a true covariant vector
(weight $W\!=\!0$), $f^*$ may be replaced with $f$, if desired. Thus are we led to
various alternate forms for $L_Q$, e.g.:
\ea{
    L_Q & = &   {\hbar^2 \over 8m}\, \sof{\rho(t,{\bf x})}^{-2} \!
       \g^{ij}\, \p_i\!\sof{\rho(t,{\bf x})} \p_j\!\sof{\rho(t,{\bf x})} \nonumber \\
       & = &  {\hbar^2 \over 8m} { \sof{\P^i f({\bf C})} \sof{\P_i f({\bf C})}
        \over  f({\bf C})^{2}} \label{LQalt}
 }

The expressions above are all manifestly covariant with respect to arbitrary
coordinate transformations of the $C^i$ that do not depend (even parametrically)
on $t$---i.e., transformations of the form $C^i \ra C^{i'}= C^{i'}({\bf C})$. Thus, for 
example, the uniformizing choice ${\bf C} = {\bf P}$ gives rise to these same expressions, 
but with ${\bf C}$ replaced with ${\bf P}$, and $f({\bf C})$ replaced with $1$. It is useful, 
however, to extend the range of coordinate transformations to include those that {\em do} 
depend on $t$, at least in the parametric sense indicated above.  We refer to such 
general coordinates as $X^j$, and see that they include both $C^i$ and $x^l$ as 
special cases. In $X^j$ coordinates, the generalized scalar probability 
density (weight $W\!=\!-1$) is denoted $\rho(t, {\bf X}) = R(t,{\bf X})^2$.  
We can therefore write the general form of $L_Q$ as
\ea{
       L_Q &  = &   
         {\hbar^2 \over 8m} { \sof{\P^j \rho(t,{\bf X})} \sof{\P_j \rho(t,{\bf X})} \over  \rho(t,{\bf X})^{2}}
         \nonumber \\ 
         &  = & 
        {\hbar^2 \over 2m} { \sof{\P^j R(t,{\bf X})} \sof{\P_j R(t,{\bf X})} \over  R(t,{\bf X})^{2}}.        
        \label{LQalt2}
}

Consider the second form in \eq{LQalt2} above, for the specific choice $X^j  = x^l$. The 
covariant derivatives become ordinary partial derivatives (gradients), 
even though $R$ is a scalar density of nonzero weight ($W\!\!=\!\!-1/2$).
The resultant
\eb
	\L_Q = (\hbar^2 / 2m) [\grad R(t,{\bf x})]\! \cdot\! [\grad R(t,{\bf x})]
	\label{KGlike}
\ee
closely resembles the gradient contribution to the KG Lagrangian---but with 
the partial derivative components restricted to the simultaneity submanifolds  
(Sec.~\ref{KG}). 
  
Another important dynamical quantity is the quantum potential, $Q$, which,
in generally covariant form, is given as
\eb
	Q =  - {\hbar^2 \over 2m} {\P^j \P_j R(t,{\bf X}) \over  R(t,{\bf X})}.
	 \label{Qdef}
\ee
The quantum potential is a true scalar invariant quantity.
For $X^j = x^l$, $\P^j \P_j$ becomes the usual Laplacian, and so \eq{Qdef} reduces to the 
familiar expression from Bohmian mechanics.\cite{bohm52a,bohm52b,holland,wyatt}  
In this form, $2\,R\,Q$ is the contribution
to the Euler-Lagrange PDE that results from $\L_Q$, treating $R$ (rather than ${\bf x}$)
as the dependent field. This matches the corresponding KG PDE contribution---but 
again---restricted to the simultaneity submanifolds (Sec.~\ref{KG}).

There is another interesting relation between $Q$ and $L_Q$, that arises in a $X^j = C^i$ 
coordinate representation.  Here, the quantity $R(t,{\bf X})$ in \eq{Qdef} becomes 
$f({\bf C})^{1/2}$.  However, it is more convenient to replace $f$ with the scalar 
invariant $f^*$, as discussed above.  Finally, since the covariant Laplacian 
$\P^j \P_j= \P_j \P^j$ is now being applied to a true scalar invariant, it can be expressed
explicitly  in terms of  ordinary partial derivatives using the Laplace-Beltrami form. 
The result is:
\ea{
     Q[x^l_{C^i}, x^l_{C^iC^j},x^l_{C^iC^jC^k}] &  = &   - {\hbar^2 \over 2m} \of{
     {1 \over \g^{1/4} f^{1/2}}} \times \qquad\quad \label{Qdef2} \\
      & & \p_i \!\sof{\g^{1/2} \g^{ij} \,\p_j\!\of{f^{1/2} \over \g^{1/4}}}, \nonumber
}
where $x^l_{C^i} = \p_i x^l = \p x^l / \p C^i$, etc.

Equation~(\ref{Qdef2}) is the trajectory-based form of $Q$, which is appropriate 
when solving for the solution trajectory ensemble, $x^l(t, {\bf C})$. Since
$\tilde \g$ depends explicitly on $x^l_{C^i}$, as discussed, the 
quantity $Q$ is third order in the $C^i$ derivatives of $x^l$. The
relation to $L_Q$ is that one may substitute the $L_Q$ form used in
the Lagrangian of \eq{nonrelLagrange}, with $L_Q = Q$, without altering 
the resultant  Euler-Lagrange PDE for $x^l(t, {\bf C})$.  This change therefore 
amounts to a change of gauge---with the $L_Q=Q$ choice offering certain 
advantages, despite the higher order, as discussed in previous 
work.\cite{poirier10nowave,poirier11nowaveCCP6,poirier12nowaveJCP} 
In any  case, it  can be shown that \eq{Qdef2} is equivalent to the 
${J^{l}}_{\! i}$ and ${K^{i}}_{ \! l}$-based expressions for $Q$ that have been
derived previously.

Whether or not $Q$ is used in the Lagrangian, it invariably plays a direct role 
in the resultant Euler-Lagrange PDE for $x^l(t, {\bf C})$. Specifically,
the covariant quantum force vector is the (negative) gradient of the quantum 
potential---i.e.,  
\eb
	f^Q_j = - \P_j Q, 
	\label{nonrelfQ}
\ee 
in generally covariant form.  The quantum force enters the 
PDE in the manner expected:
\ea{
	m {\ddot x}^l +  \d^{lm}\prt{V({\bf x})}{x^m}  &  =  &      f^l_Q ,\qquad {\rm where}
	       \label{nonrelPDE} \\
	  f^l_Q  & = & {\hbar^2 \over 2m}\,
	        \d^{lm} {K^{k}}_{\!m}\,  \p_k\! \left ( {1 \over \g^{1/4} f^{1/2}} \times \right . \nonumber  \\
	              & & \left . \p_i 
	        \!\sof{\g^{1/2} \g^{ij} \,\p_j\!\of{f^{1/2} \over \g^{1/4}}} \right ),  \nonumber
}
and ${K^{k}}_{\! m} = \p C^k / \p x^m$.  

The Euler-Lagrange PDE for  $x^l(t, {\bf C})$---i.e., \eq{nonrelPDE}---is
second order in time ($t$) and fourth order in space $({\bf C})$.
Note, however, that no time derivatives enter into the expressions
for the quantum (and classical) forces. At a given point $p$, therefore,
the $f^l_Q$ vector belongs to the orthogonal subspace of $T_p$,  
and is otherwise only constructed from quantities that also belong 
to the simultaneity submanifold associated with $p$. Postulate 2 is 
therefore satisfied, for the trajectory-based formulation of 
non-relativistic quantum mechanics.


\section{Relativistic Quantum Dynamics}
\label{relativistic}

\subsection{Relativistic quantum dynamical equation of motion}
\label{relPDE}

We have taken considerable lengths to lay the groundwork for our
main objective: an exact, self-contained, trajectory-ensemble-based 
PDE, describing the relativistic quantum dynamics of a spin-zero free particle.
This foundational effort has been well spent, in that it gives rise to an 
essentially unique relativistic quantum formulation, requiring only the barest  
minimum of  assumptions. 

Let us start with the relativistic quantum potential, $Q$.  From previous
non-relativistic work, it is known that the allowed trajectory-based form 
for $Q$ is essentially unique.\cite{poirier10nowave,poirier11nowaveCCP6,poirier12nowaveJCP}  
The same basic form should be required in the relativistic context---although here, 
there might in principle be some 
question as to which metric tensor (i.e., 3d or 4d) and/or probability density
quantity should be employed.  Postulate 2, along with the 
broader discussion of Sec.~\ref{classical}, completely alleviates this ambiguity.  In particular,
since $Q$ directly gives rise to quantum forces, it 
must be constructed on the relativistic simultaneity submanifold, using the 
3d relativistic spatial metric tensor $\tilde \g$ of \eq{gblock} and Sec.~\ref{general}.  
(More technically---and for completely general
coordinates $X^\mu$ and metric tensors $\tilde g$---the spatial metric tensor $\tilde \g$
that is used for this purpose is the induced metric tensor on the simultaneity 
submanifold,\cite{carroll,weinberg} i.e. the pullback of $\tilde g$).  
Likewise, the relevant probability quantity is 
the spatial scalar probability density, $f({\bf C})$. As discussed in Sec.~\ref{electromagnetism}, 
$f({\bf C})$ ``lives'' on the simultaneity
submanifolds, and, is therefore unaffected by reparametrizations, $\l \ra \l'(\l)$, 
of the ensemble time coordinate, $\l$.  

Without further ado, then, we posit a {\em relativistic quantum potential} of the 
form of \eq{Qdef2}, i.e. 
\ea{
     Q[x^\a_{C^i}, x^\a_{C^iC^j},x^\a_{C^iC^jC^k}] &  = &   - {\hbar^2 \over 2m} \of{
     {1 \over \g^{1/4} f^{1/2}}} \times \qquad\quad \label{Qdef3} \\
      & & \p_i \!\sof{\g^{1/2} \g^{ij} \,\p_j\!\of{f^{1/2} \over \g^{1/4}}}. \nonumber
}
One important difference between \eqs{Qdef2}{Qdef3} is that in the latter,
$t=x^0/c$ is a coordinate, rather than a parameter.  Thus, for example, 
$\p x^0 / \p C^i$ appears explicitly in the expression for $\tilde \g$, in the 
relativistic case (along with $\p x^l / \p C^i$, of course).   Another important 
difference is that the $\p_i = \p / \p C^i$ partial derivatives occur at constant $\l$, 
rather than at constant $t$.   We might have expected the relativistic partial 
derivatives to be at constant $\t$; the fact that these occur at constant $\l$, instead,
is very significant.  In any event, the relativistic quantum potential as defined
by \eq{Qdef3} is invariant with respect to reparametrizations of
both $\l$ and ${\bf C}$.

Another significant feature of \eq{Qdef3} is that the explicit partial derivatives,
$\p_i$, extend over only the spatial, $C^i$ components of the natural coordinates
[$X^\mu$ from \eq{natcoord}].  The lack of a timelike, $\l$ contribution in 
effect implies that the 4d Laplace-Beltrami operator has been projected onto the simultaneity 
submanifold---a necessary consequence of Postulate 2, of course. It is possible 
to develop a completely general covariant expression for $Q$, in terms of completely 
arbitrary spacetime coordinates.  In this case, however, the relativistic analog of 
\eq{Qdef} must be substantially modified, to explicitly incorporate the requisite 
projection tensors  associated with the pullback of $\tilde g$.\cite{carroll,weinberg} 
We do not find it profitable to do so here. 

The relativistic quantum force is obtained as the (negative) covariant gradient
of the quantum potential scalar invariant of \eq{Qdef3}.  Again, Postulate 2 requires that only the 
gradient components corresponding to the $C^i$ coordinates be considered---i.e.,
the gradient is restricted to the simultaneity submanifold.   In a completely general 
coordinate system, each of the gradient components of the force vector
must be explicitly projected onto the submanifold, thus modifying the form of 
\eq{nonrelfQ}.  In the $x^\a$ inertial coordinates, for example, this projection is effected 
using the twelve $\mu=i$ components of the inverse Jacobi matrix---i.e., 
the ${K^{i}}_{\!\a} = \p C^i / \p x^\a$. The inertial components of the 
relativistic quantum force vector,  $f^\a_Q$, are thus given explicitly by  
\ea{
                 f^\a_Q & \!=\! &   - \eta^{\a\b}  {K^{k}}_{\!\b} \, \p_k \,Q  \label{relfQ} \\
                 &\! =\! &  
                 {\hbar^2 \over 2m} \eta^{\a\b}  {K^{k}}_{\!\b}  \p_k \!
	       \of{{1 \over \g^{1/4} f^{1/2}} \p_i 
	        \!\sof{\g^{1/2} \g^{ij} \,\p_j\!\of{f^{1/2} \over \g^{1/4}}}}. \nonumber
}

As discussed in Sec.~\ref{classical}, the time coordinate associated with the dynamical
evolution is $\t$, rather than $\l$. For the relativistic quantum free particle,
the only dynamical force is the quantum force, defined by \eq{relfQ}. We 
therefore expect the {\em relativistic quantum equation of motion to be given
by \eq{reldyn}, with $f^\a = f^\a_Q$}. This equation thus constitutes
a new (to our knowledge) dynamical prediction, that can in principle be 
tested against experiment.  It can be interpreted as a PDE, but one that is not
equivalent to the KG PDE (Sec.~\ref{KG}).  The proposed dynamical law is 
unambiguous, explicit, and exact.  However, in its present form, it has a
``mixed'' character, whereby $\l$ is the time coordinate used in the determination
of $Q$ and $f^\a_Q$, but $\t$ is what is used in the dynamical law.  Later, 
we will derive a single, consistent form, expressed solely in terms of $\l$.  
First, however, we will demonstrate that the proposed dynamical law is in
fact the Euler-Lagrange equation that results from extremizing the corresponding
relativistic quantum action.

\subsection{Extremization of the relativistic quantum action}
\label{relativistic-action}

The relativistic quantum dynamical law proposed in  Sec.~\ref{relPDE}
 [i.e. \eqs{reldyn}{relfQ}]
should be derivable as the Euler-Lagrange PDE obtained from the 
extremization of some suitable action quantity.  Moreover, from a perspective
of general covariance,  this action should be a true scalar invariant, obtained
as an integral over $d^4 X$, for arbitrary coordinates, $X^\mu$. In analogy
with \eq{nonrelaction}, therefore, we expect
\ea{
	S &  = & \int \! \L[x^\a, x^\a_{X^\mu},\ldots]  \, d^4 X  \nonumber \\
	    &  = & \int \! L[x^\a, x^\a_{X^\mu},\ldots] \,  {\rho(X^\mu) \over c} \, d^4 X,
	\label{relaction}
}
where $\rho(X^\mu)$ is the  scalar probability density of \eq{rhogen},
and the Lagrangian, $L$, is a true scalar invariant with units of energy.

If  $X^\mu$ is taken to be a set of natural coordinates, then \eq{rhonat} 
implies that \eq{relaction} receives an extra factor of  $\sqrt{-g_{00}} = \t_\l$ 
in comparison with \eq{nonrelaction}, where $\t_\l$ is defined as in 
Sec.~\ref{electromagnetism}.  This fundamental difference from the non-relativistic
case is associated with the fact that it is $\t$ rather than $\l$ that is the dynamical evolution 
coordinate in the relativistic context.  In any case, from the discussion in 
Sec.~\ref{electromagnetism} [e.g., 
\eqs{freelam}{EMlam}] and Sec.~\ref{nonrel} [\eq{nonrelLagrange}] the explicit form
of the relativistic quantum action, in natural coordinates, is as follows:
\ea{
      S & = &  \int \!  \sof{(-mc^2) \sqrt{- {1 \over c^2}\, \eta_{\a\b}\, x^\a_\t \, x^\b_\t }
                                -L_Q}\!  \t_\l\, f({\bf C})   \, d^3C\, d\l \nonumber   \\
          & = &  \int (-mc^2) \!\left [  \sqrt{- {1 \over c^2}\, \eta_{\a\b}\, x^\a_\l \, 
                         x^\b_\l}\,\,  + \t_\l\, {1\over 8} \of{{\hbar \over mc}}^2  
                          \times \right .   \nonumber \\
          &   &    \left .   \sof{f({\bf C}) \over \g^{1/2}}^{-2}\!\!\!\! \g^{ij} \p_i\!\!\sof{f({\bf C}) \over \g^{1/2}}
                       \! \p_j\!\!\sof{f({\bf C}) \over \g^{1/2}} \right ]
                        \!\! f({\bf C})    d^3C d\l   \label{relaction2} }

Since $\t_\l$ depends explicitly on the $x^\a_{X^\mu}$
and implicitly on the $X^\mu$, its presence in the Lagrangian density definitely
complicates matters, vis-\`a-vis the resultant Euler-Lagrange PDE. Matters
are further complicated by the fact that the $\l$ coordinate is not defined {\em a priori},
as discussed in Sec.~\ref{electromagnetism}. We therefore anticipate difficulties, or at 
least complications, in any Euler-Lagrange derivation based on natural coordinates, 
$X^\mu = (c\,\l, {\bf C})$. The most expedient way around this difficulty is to adopt a kind 
of hybrid approach, involving a change of coordinates,
 $X^\mu \ra Y^\nu(X^\mu) = (c \,\tau(X^\mu), {\bf C})$, where the time coordinate for 
 the new frame, $Y^\nu$, has been changed from  $\l$ to $\t$. 

An advantage of working in the $Y^\nu$ frame is that the time coordinate $\t$ 
is now well determined. Overall, however, the above change of coordinates does not 
necessarily  constitute an improvement; for example, the new metric tensor, 
$\tilde g$, is no longer block diagonal.  Also, the restriction to the simultaneity 
submanifolds of quantities such as $L_Q$ and $Q$ would require explicit use of 
projection tensors. On the other hand, we recall \eq{taudef}, which allows us to 
redefine the proper time coordinate, $\t$, by applying an arbitrary, path-dependent
shift, $\Delta ({\bf C})$.  This enables us to fix the particular choice of $\t$, such that 
{\em one} of its contours, at least,  coincides exactly with a simultaneity submanifold.  

Let us, then, hereby choose $\Delta ({\bf C})$ such that $\l\!=\!0$ corresponds to $\t\!=\!0$.  
As a consequence, for all  points $p$ belonging to the specific simultaneity 
submanifold specified by $\l=\t=0$,
the partial derivatives $\p_i | \l$ and $\p_i | \t$ are the same, and so the latter can
replace the former in the expressions for $L_Q$, $Q$, and $f^i_Q$.  Along this
simultaneity submanifold, therefore, the Lagrangian in $Y^\nu$ becomes
\ea{
      L & = &   (-mc^2) \!\left [  \sqrt{- {1 \over c^2}\, 
                         \eta_{\a\b}\, {dx^\a \over d \t} \, {dx^\b \over d\t}}\,\,  + \right .  
                          \label{tauaction} \\
       &   & 
                         \left .   {1\over 8} \of{{\hbar \over mc}}^2 \sof{f({\bf C}) \over \g^{1/2}}^{-2} \!\!
                         \g^{ij}\, \p_i\!\!\sof{f({\bf C}) \over \g^{1/2}} \p_j\!\!\sof{f({\bf C}) \over \g^{1/2}} \right ]
                           \nonumber 
}
where the $Y^i\!\! = \!\!C^i$ partial derivatives, $\p_i$---explicit in \eq{tauaction} above, 
and implicit in the definition of the $x^\a_{Y^\nu}$ quantities used to construct the
spatial metric tensor $\tilde \g$---are taken at constant $\t$, rather than at constant $\l$. 

The advantage of \eq{tauaction} is that it is now very straightforward to derive
the corresponding Euler-Lagrange PDE---at least as this is restricted to the 
$\l=\t=0$ simultaneity submanifold.  The result is indeed the dynamical law of 
Sec.~\ref{relPDE}---i.e., \eqs{reldyn}{relfQ}, with $f^\a = f^\a_Q$.  Note also that these
equations have no explicit dependence on the $\t$ coordinate itself---only on 
the differential, $d\t$, which is invariant with respect to transformations of 
the form of \eq{taudef}.  This is significant, because it implies that the dynamical
law must hold not only for the $\l=\t=0$ simultaneity submanifold, but for 
{\em all} simultaneity submanifolds---and therefore, across the entire 
spacetime manifold, $M$.

\subsection{Conversion to natural coordinates, 
and  ``dynamical'' ensemble proper time, $\T$}
\label{conversion}

Our present theory encompasses a hybrid dynamical equation,
wherein $\t$ is the time evolution coordinate associated with the 
quantum force vector $f^\a = f^\a_Q$, but  the latter is itself obtained 
at constant values of the ensemble time coordinate, $\l$.  (We henceforth drop the 
$Q$ subscript on the quantum force vector, $f^\a$). In practical terms, 
such an equation is not so useful for actually effecting propagation of the 
solution  trajectory ensemble, $x^\a(X^\mu)$.  We therefore seek to convert
the dynamical equation to a more consistent form involving only 
the natural coordinates, $X^\mu = (c\,\l, {\bf C})$, that can 
be directly integrated with respect to $\l$, and without recourse to $\t$.  

Since the equation of motion is second order in time, the quantities
to be propagated may be taken to be $x^\a({\bf C})$ and $U^\a({\bf C})$. 
We seek new time evolution equations for the derivatives of these
quantities with respect to $\l$.  From \eqs{rhonat}{genflux}, and also
\eq{reldyn}, these are found to be:
\ea{
	{d x^\a \over d\l}  & = &  c \left . \prt {x^\a} {X^0} \right |_{\bf C}  = \t_\l \, U^\a \nonumber \\
	\quad {d U^\a \over d\l} & = &   c  \left . \prt {U^\a} {X^0} \right |_{\bf C}  =  \t_\l \,{f^\a \over m}
	\label{relquantdyn}
}
 
We therefore have what we need in \eq{relquantdyn}---provided that
an explicit expression for $\t_\l$ can be obtained.  In general, this quantity varies across
spacetime, and even across a simultaneity submanifold.  In addition,
it will change under a reparametrization of the $\l\ra \l'(\l)$ form. 
Also, in the special case of quantum inertial motion---provisionally defined 
via $f^\a=0$ for all $x^\a$ (a more precise definition will soon be introduced)---an
arbitrary  $\l$ should satisfy $\t_\l(X^\mu) = h(\l)$, for some function  
$h(\l)$. In other words, for the inertial case, $\t_\l$ should not vary across a 
given simultaneity submanifold.   Choosing $\l$ to be an ensemble proper time, 
$\T$, one has $h(\T)=1$ for the inertial case, by definition. In any case,  constant $\t_\l$ 
across a given simultaneity submanifold implies $f^\a=0$, and so conversely, we expect
nonzero quantum forces to give rise to local variations in $\t_\l$.
  
What precise relationship should one expect to see between  $\t_\l$ variations 
and the quantum force? Consider that a $\l$ reparametrization has the effect of 
rescaling $\t_\l$ uniformly across a given simultaneity submanifold.  
As discussed, such a reparametrization
has no effect on $f^\a$. Therefore, regardless of how $\l$ defined,  it is the
{\em relative} changes in $\t_\l$---i.e., $\p_i (\t_\l) / \t_\l$---that should appear
in the quantum force relation. 

In any event, an explicit form for this relation
may be obtained upon transforming \eq{relquantdyn} to a natural coordinate 
frame.  Treating $\t$ in this context 
as a parameter, or scalar field, we have already seen that both $x^\a$ and $U^\a$
are true four-vectors.  The covariant derivative of the $U^\a$ vector field, i.e.  
\eb
         {T_{\b}}^{ \a} = \p_\b U^\a \qquad ; \qquad {T_{\nu}}^{ \mu} = \P_\nu U^\mu,
\ee
is also a  true tensor invariant, where the second equation above is the generally 
covariant form.  Acting on an arbitrary 
displacement vector, $d x^\b$, this tensor yields the corresponding change
in the velocity four-vector, $dU^\a$.  To compute the change in $U^\a$ 
along the direction of motion itself, ${T_{\b}}^{ \a}$ should be contracted with
$U^\b / c$. Upon dividing both sides by $\t_\l$, the second part of  \eq{relquantdyn} 
thus becomes
\ea{
         {d {U^\a} \over d \t }  & = &  U^\b\, {T_{\b}}^{ \a}    =     {f^\a \over m} \label{natdyn} \\
	U^\nu\, {T_{\nu}}^{ \mu} &  =  &  U^\nu\, \P_\nu U^\mu =  
	 U^\nu \of{ \p_\nu U^\mu  + \G^\mu_{\nu\s} U^\s}   = {f^\mu \over m}, \nonumber 
}
where the second equation above is the generally covariant form.

From \eqs{fourv}{Vdef}, the natural coordinate form of $U^\mu$ is easily found to be
\eb
	U^\mu = \of{{c \over \t_\l}, 0,0,0}.
	\label{Unat}
\ee
From \eq{Unat}, it is also easy to verify the natural coordinate generalization 
of the normalization condition, \eq{fourvcond}.  Likewise, the orthogonality 
of $U^\a$ and $f^\b$---i.e., \eq{relfcond}---must also hold in the 
natural coordinate frame:
\eb
          g_{\mu\nu}\, U^\mu f^\nu = 0
	\label{natrelfcond}
\ee
From \eqs{Unat}{natrelfcond}, and the block-diagonal nature of $\tilde g$
[\eq{gblock}], it follows that 
\eb
	f^\mu = (0, {\bf f}) =  (0, f^i) = (0, f^1,f^2,f^3),
	\label{fnat}
\ee
in a natural coordinate representation. The fact that $f^0\!=\!0$ is 
consistent with the requirement that $f^\mu$ live on the simultaneity
submanifold.  However, the metric ensures that  even though the 
$f^\mu$ vector is always ``horizontal,'' the $U^\mu$ vector is 
nevertheless always ``vertical''	 (Fig.~\ref{fig-natural}).

The results of the previous paragraph must be reconciled with \eq{natdyn}.
In natural coordinates,
\eb
	\G^0_{00} = {1\over 2}\, g^{00}\, \p_0 g_{00} = {1 \over 2c} \of{{1 \over \t_\l^2}}
	 \prt {\t_\l^2}{\l}.  \label{G000}
\ee
Substituting \eqs{Unat}{G000} into \eq{natdyn} thus yields $f^0=0$, as expected.
It is the expression for the spatial components of force, however, that yields the 
desired $\t_\l$ relation.  Here, the result
\eb
	\G^i_{00} = -{1 \over 2} \,\g^{ij}\, \p_j g_{00}  = 
	                      {1 \over 2} \,\g^{ij}\, \p_j \t_\l^2 \label{G100}
\ee
gives rise to
\eb
	{ f^i \over m} = c^2\, \g^{ij} \of{{\p_j \t_\l} \over \t_\l},
         \label{fiform}
\ee
which is exactly of the form desired, as discussed above. 

Equation~(\ref{fiform}) above may be rewritten in more general
and suggestive fashion as
\eb
	f^i = m c^2 \,\g^{ij}\, \p_j\!\sof{\log\!\of{A(\l) \t_\l}},
	\label{fiform2}
\ee
where $A(\l)$ is constant across a given simultaneity submanifold,
but depends arbitrarily on $\l$. Clearly, the function $A(\l)$ defines
a particular $\l$ parametrization; it has units of $\l$ over time.  The
most natural choice is simply $A(\l)=1$.  With this choice, $\l$
has units of time, and satisfies $\t_\l = 1$ in the case of inertial 
motion.  It is therefore an ensemble proper time, $\l = \T$.  We call
this choice the ``dynamical ensemble proper time'';  henceforth, it is
the only choice for $\T$ that will be considered.  

It is necessary to demonstrate consistency of \eq{fiform2} across
all values of $i$.  This follows easily from the fact that $f^i$ is the
(spatial) gradient of the quantum potential [\eq{nonrelfQ}], thus 
enabling a direct comparison between $\t_\l$ and $Q$. Choosing
$\l=\T$, this leads to the fundamental result,
\eb
	{d\t \over d\T} = \t_\T = \exp\!\sof{-{Q \over mc^2}} .
	\label{tldef}
\ee
The importance of \eq{tldef} cannot be overstated. In addition
to providing exactly the $\t_\l=\t_\T$ form needed to propagate \eq{relquantdyn},
this expression is imbued with fundamental physical significance. 
Specifically, it states that {\em the quantum potential induces a
time dilation or contraction for individual trajectories}, much like 
gravitational time dilation. Also like the case of gravity---but unlike most other
physical forces---the absolute quantum potential matters.
Thus, under the transformation, $Q \ra Q + {\rm const}$,  there is a
corresponding change in $\T$---a constant rescaling, in fact---even though
the quantum forces and accelerations are unaltered.  
Put another way, \eq{tldef} requires that quantum inertial motion 
satisfy a more restrictive condition than stated above; 
specifically, we must have $Q\!=\!0$ everywhere, rather than just $f^i\!=\!0$.

An important difference from the case of gravity is that  the quantum
potential $Q$ may be either positive {\em or} negative.  In other words, the
passage of proper time for a given quantum trajectory may 
be slower {\em or faster} than that for the whole ensemble,
depending on the sign of $Q$.  There is also a curious inverse behavior, 
whereby lower (i.e. more negative) values of the quantum potential 
lead to {\em faster} evolution of $\t$ (relative to $\T$)---exactly the opposite 
of the gravitational case.  More specifically, in  the weak field limit, one has
\eb
         -\of{1 + 2 { m \Phi \over m c^2}} \approx g_{00}
         \approx -\of{1- 2 {Q \over m c^2}} 
         \label{grav},
\ee
where the left hand side of \eq{grav} is for GR, and the right hand side 
is for the present quantum theory. Thus, it is $-Q$, rather than $Q$ itself,
that plays the role of the gravitational potential energy, $m \Phi$. 

On the other hand,  the above behavior makes perfect sense, from 
another perspective. In standard Bohmian mechanics,
regions of space where $Q$ is positive may be (loosely) identified as
classically allowed regions, whereas negative $Q$ generally denotes
classically forbidden regions into which quantum tunneling can 
occur.\cite{holland,wyatt}
In the classically allowed regions, therefore, the proper time for a given trajectory
is dilated---as would be predicted by classical relativity theory.  
Quantum tunneling, on the other hand, speeds up the passage
of proper time---a phenomenon with no classical relativistic analog. 
Although the present work focuses on the flat spacetime manifolds of SR, the
features of relativistic quantum dynamics discussed above suggest an extremely 
interesting interplay between gravitational and quantum forces, in the 
context of quantum gravity. 

In any event, substitution of \eq{tldef} into \eq{relquantdyn} provides
us with our goal for this subsection---a set of dynamical equations, 
consistently expressed in terms of any choice of  natural coordinates, 
$X^\mu = (c\,\l, {\bf C})$, that can be directly integrated over $\l$.
[To obtain the general-$\l$ form from the $\l\!=\!\T$ form, one simply 
introduces  a factor of $A(\l)$ into the exponent of \eq{tldef}.]  These equations, for
$dx^\a / d\l$ and $dU^\a/d\l$, are first  order in $\l$ and fourth order 
in ${\bf C}$, with all quantities on the right hand sides of \eq{relquantdyn} 
obtained from  a single simultaneity submanifold [i.e., from $f({\bf C})$, $x^\a({\bf C})$, 
and various $\p_i$ partial derivatives,  all at fixed $\l$].  
We also find it convenient to express this
dynamical law as a second order (in time) PDE, using the specific choice of 
natural coordinates, $X^\mu = (c \,\T, {\bf P})$,
where $\T$ is the dynamical ensemble proper time, and 
${\bf P}$ are the uniformizing coordinates of Sec.~\ref{uniformizing}. The result is:
\ea{
     \prtsq{x^\a}{\T} &  =  &  \exp\!\sof{-{2Q \over mc^2}} {f^\a \over m} -
             \of{ {1 \over mc^2}} \prt{Q}{\T}\, \prt{x^\a}{\T}  \nonumber \\
      {\rm where} \quad  Q &  = &  - {\hbar^2 \over 2m} \,
     \g^{-1/4} \,\p_i \!\sof{\g^{1/2}\, \g^{ij} \,\p_j \g^{-1/4}} \nonumber \\
     {\rm and} \quad f^\a &  =  & - \eta^{\a\b}  \,\prt{C^i}{x^\b} \, \p_i Q = 
                                                     -  \prt{x^\a}{C^i} \,\g^{ij} \, \p_j Q \label{finalPDE} 
}

In practice, the last form of the last line above may be more convenient than the middle
form,  because it involves explicit partial derivatives of the $x^\a$  rather than
of the $X^\mu$ (i.e., the elements of the Jacobian matrix ${J ^{\a}}_{\!\mu}$,
rather than the inverse Jacobian ${K ^{\mu}}_{\!\a}$). It can be easily 
derived, simply by raising the index of the force vector prior to transforming
from natural to inertial coordinates [e.g., via \eq{fiform}], and exploiting
 \eqs{gblock}{fnat}.  
It can also be shown that the various constraints imposed on $U^\a$, $f^\a$,
etc., as discussed previously in this document, are all preserved under 
the evolution of \eq{finalPDE}.

\subsection{Classical, inertial,  and non-relativistic limits}
\label{limits}

As a necessary check on the viability of our candidate relativistic
quantum dynamical law, as presented in Secs.~\ref{relPDE} 
and~\ref{conversion}, we must consider various  asymptotic limits 
for which the correct behavior is known.  
Two obvious 
limits that must be considered are the relativistic classical limit, and the 
non-relativistic quantum limit. Fortunately, the trajectory-based formulation 
adopted here renders both of these comparisons very straightforward.
In addition to these two limits, one should also consider the limit of quantum inertial 
motion, alluded to several times previously in this document.

The relativistic classical limit is the most straightforward. Consider the
full, relativistic quantum action, as presented in \eq{relaction2}. 
We define the ``relativistic classical limit'' of this action as what ensues
by setting $\hbar=0$. Of course, the second, $L_Q$ term vanishes 
completely in this case. One is indeed left with the relativistic classical
action of \eq{freetau}, except that it is integrated by $f({\bf C})\, d^3C$.   
This integration is immaterial, however, because the relativistic classical
Lagrangian density---being truly classical---does not depend explicitly
or implicitly on the $x^\a_{C^i}$.  Thus, in the relativistic classical limit,
each trajectory of  the solution ensemble propagates completely 
independently of the others, and in accord with the relativistic classical 
equation of motion, \eq{reldyn}.  This behavior is completely as 
expected---being analogous, e.g., to the Hamilton-Jacobi trajectory
ensemble in the non-relativistic classical theory.  

Note that in both relativistic and non-relativistic contexts, individual trajectories 
may cross each other, in the classical limit.   For semiclassical and related 
approximations to non-relativistic quantum mechanics that operate in the
classical limit, trajectory crossings lead to well-known headaches
such as caustics, divergent probability densities, multivalued field functions, 
etc.  For such approximations, also, it is well-known that  ``as $\hbar$ goes 
to zero'' is a  rather imprecise phrase, whose specific mathematical interpretation 
 must be handled very delicately.  In the trajectory-based formulation 
 of non-relativistic quantum mechanics,
 for example, setting $\hbar=0$ leads to the semiclassical or Hamilton-Jacobi 
 behavior described above (e.g., with crossing trajectories)---whereas setting $\hbar$
 to {\em any} finite value, no matter how small, leads to a qualitatively 
 very different ensemble with no trajectory crossings.\cite{holland,wyatt}  
 
The same is true for the present relativistic quantum formulation.  Thus, 
it is necessary to distinguish the case where $\hbar\!=\!0$ identically 
(considered above) from that where $\hbar$ {\em approaches} zero 
(in a certain well-defined sense given below).  Simply put, in the latter case, 
there is still {\em correlation} across the trajectories in the ensemble, even 
though quantum {\em dynamical} effects  become arbitrarily small. It is easy to
 show that this limit corresponds to the
special case of quantum inertial motion, for which 
$X^\mu = (c \T, {\bf P'}) \ra x^{\a'}$---with ${\bf P'}$ a rescaled version of ${\bf P}$,
and $x^{\a'}$ some inertial coordinate frame [see \eq{LT}].

More precisely, quantum inertial motion is defined via the limit,
\eb
	{L_Q \over mc^2} = {1\over 8} \of{{\hbar \over mc}}^2
       \sof{{\p^i \rho(t,{\bf x}) \over \rho(t,{\bf x})}}  \sof{{\p_i \rho(t,{\bf x}) \over \rho(t,{\bf x})}}      
       \ra 0 
       \label{inertiallimit}
\ee
The physical significance of \eq{inertiallimit} is clear; it says that quantum 
inertial behavior emerges when the length scale of the probability distribution 
is extremely large compared to the Compton wavelength, $\hbar/mc$.  Thus, 
$\rho(t, {\bf x})$ approaches ``infinite broadness,'' or  effectively uniform behavior. 
There is an obvious analogy with the Einstein Equivalence Principle, except
that in the relativistic quantum context, ``sufficiently small regions of spacetime'' must still
be larger than the Compton wavelength. 

One can equally well define the quantum inertial limit via $(Q/mc^2) \ra 0$, 
from which \eq{tldef} leads to $\t_\T \ra 1$ or $\T \ra \t$---a necessary condition
for quantum inertial motion. Because the quantum potential approaches zero,
the same must be true of the quantum force vector, $f^i_Q$, implying that 
the quantum trajectories follow nearly perfect straight-line orbits, as in 
the classical limit case.  However, unlike the classical limit case, the trajectories
within a given ensemble are all parallel, in accord with \eq{inertialdef}.  This assertion is
easy to prove by contradiction. Suppose that two trajectories in the ensemble
are {\em not} parallel; they must cross at some point $p$ in $M$. This violates
the presumed ensemble conditions of Sec.~\ref{basic}, but according to Postulate 1,
the probability density $\rho$ must diverge at $p$, which also violates \eq{inertiallimit}.
Thus, in a global sense at least (i.e., across all $M$), quantum inertial motion 
implies \eq{inertialdef}.

Finally, we consider the non-relativistic quantum limit. Here, we may expect that 
\eqs{relaction2}{finalPDE} will reduce to the corresponding expressions
for the non-relativistic trajectory-based formulation presented in 
Sec.~\ref{nonrel}---but with $V({\bf x}) = 0$, as is appropriate for a free particle.  First, we define
the relativistic parameter $\b$ (not to be confused with the index $\b$) 
in the usual fashion, as the speed of the particle in units of $c$:
\eb
	\b = {\sqrt{U^l\, U_l} \over U^0} = \sqrt{{{\dot x}^l \over c} \, {{\dot x}_l \over c}}
		\label{betadef}
\ee
where $U^\a = (U^0, U^l)$ and ${\dot x}^l = dx^l / dt$. By definition, 
$0\le \b < 1$, with $\b\ra0$ taken to be the non-relativistic limit. 

To address this limit, it is therefore appropriate to expand all quantities in \eq{relaction2}
in powers of $\b$, in which context it is convenient to consider
the classical (first term) and quantum (second term) contributions separately. 
The classical contribution has no explicit dependence on ${\bf C}$, nor on its
partial derivatives; we may therefore take $\l$ to be any trajectory parametrization,
with the choice $\l = t$ a particularly convenient one. This yields  
\ea{
      \int (-mc^2)  \sqrt{1-{{\dot x}^l \over c} \, {{\dot x}_l \over c}}\,  f({\bf C})   \, d^3C\, dt  
      \,\,\,  \approx   \qquad\qquad\qquad  &   &   \\    \qquad\qquad \qquad
       \int \!\of{-mc^2 + {1\over 2} m\,  {\dot x}^l\, {\dot x}_l}  f({\bf C})   \, d^3C\, dt , & &
       \nonumber
}
where the right hand side above is the expansion of the left hand side,
to second order in $\b$.  The zeroth order contribution is the usual rest energy,
which---being a constant---has no effect on the dynamics, and can be ignored. 
There is no first order contribution. The second order contribution is the expected 
non-relativistic classical kinetic energy, i.e., the first term in \eq{nonrelLagrange}.

To maintain $\beta$ order consistency, we should expand the quantum  contribution to 
 \eq{relaction2} to second order in $\beta$ as well.  Let $\Delta t$ and $\Delta x$ denote 
 roughly the range of variation of these respective quantities, for a given trajectory. In the 
 relativistic limit of small $\b$, we have $\Delta t \approx \Delta \tau$, and so the quantum 
 force $f^\a$ is on the order of $m\, {\ddot x}^\a$.  This leads to
\eb
	{Q \over \Delta x} \approx m \,{\Delta x \over {\Delta t}^2}  \qquad ; \qquad
	{Q \over m c^2} \approx {{\dot x}^l \over c} \, {{\dot x}_l \over c} = \b^2,
	\label{Qexpand}
\ee     
where ``$\approx$'' is to be interpreted here as meaning ``on the order of.''  Similar arguments 
may be used to show that $Q \approx L_Q$.  Thus, the quantum contribution to \eq{relaction2}
is (implicitly) second order in $\beta$.  We need therefore only consider the lowest order
contribution in the explicit $\beta$ expansion of this quantity, which simplifies matters
considerably. 

To zeroth order in $\b$, the trajectory is at rest, and so $t \approx \t$, 
$U^0 \approx c$ and  $U^l \approx 0$.  The $\T$ and $t$ contours are
identical to zeroth order; likewise, the intersections of the $C^i$ contours 
(which define the trajectory) and the $x^l$ contours are identical. As a consequence,
 $(\p/\p C^i) |_\T \approx (\p / \p C^i) |_t$.  To lowest order, the   
${J^{\a}}_{\!\mu}$ tensor thus becomes 
\ea{
	{J^{0}}_{\!0} \approx \t_\T \quad & ; &  \quad {J^{0}}_{\!i} \approx 0 \nonumber \\
          {J^{l}}_{\!0} \approx 0 \quad & ; & \quad  \left . {J^{l}}_{\!i} \approx \prt{x^l}{C^i} \right |_t
	\label{Jzero}
}
From \eqs{gdef}{Jzero}, and the discussion in Sec.~\ref{relPDE}, 
the spatial metric tensor becomes
\eb
	\g_{ij} = \eta_{\a\b}\, {J^{\a}}_{\!i} {J^{\b}}_{\!j} 
	\approx \d_{lm} \,\left . \prt{x^l}{C^i} \right |_t \left . \prt{x^m}{C^j} \right |_t,
\ee
where the final expression above is the lowest order approximation. But
this is identical to the non-relativistic $\g_{ij}$ tensor of \eq{nonrelaction}.
Finally, from \eqs{tldef}{Qexpand}, we have $\t_\T \approx 1$ at lowest
order. Thus, $d\T \approx d\t \approx dt$, and we can replace $d\l$ in
the integration of the second term in \eq{relaction2} with $dt$.  Combining
this with the second order result for the first term as derived above, 
one obtains exactly the non-relativistic Lagrangian of 
\eq{nonrelaction} [with $V({\bf x}) = 0$].

In principle, the above Lagrangian should lead, via the usual Euler-Lagrange procedure, 
to a set of dynamical PDEs that are consistent with the $\beta\ra0$ limit of the relativistic quantum
equation of motion. Moreover, since the zeroth order (rest energy) contribution to the relativistic 
quantum Lagrangian is a constant, and since there is no first order contribution, we should need 
only consider the lowest order contribution to the relativistic quantum PDE, \eq{finalPDE}
(with $\a=l$). To lowest order, the last term in this equation is zero, because 
$\p x^l / \p\T \approx  c\, {J^{l}}_{\!0} \approx 0$. Likewise, the exponential factor 
is unity, to lowest order, and the partial derivatives with respect to $\T$ on the left hand
side may be replaced with $t$ derivatives.  The result, therefore, is indeed identical
to \eq{nonrelPDE} [again with $V({\bf x}) = 0$].

\subsection{Comparison with Klein-Gordon equation}
\label{KG}

The standard description for a massive, spin-zero, relativistic quantum particle
is, of course, that of  the famous Klein-Gordon (KG) equation.\cite{bohm,messiah,holland,debroglie56,feshbach58,carroll,aharonov69,blokhintsev,ranada80,kyprianidis85} This is a wave PDE, for which the independent variables are $x^\a$, and the dependent 
quantity is the real- or complex-valued wave field, $\Phi(x^\a)$.  The KG PDE takes the form
\eb
	\mu^2 \Phi - \p^\a \p_\a \Phi = 0, 
	\label{KGPDE} 
\ee
where $\mu$ is a positive real constant (not to be confused with the index).  
 The KG PDE is linear, and second order, in 
both the spatial coordinates, $x^l$, and the time coordinate, $x^0$. It can be derived 
from a simple KG Lagrangian density, involving only the first order derivative quantities, 
$\p_\a \Phi=\Phi_{x^\a}$, as well as the wave field, $\Phi$, itself:
\eb
	\L[\Phi, \p_\a \Phi] = {1 \over 2} \sof{ \mu^2 \Phi^2 +(\p^\a \Phi)(\p_\a \Phi)}
	\label{KGL}
\ee   
The solution wave field---i.e., a $\Phi(x^\a)$ that satisfies \eq{KGPDE}---extremizes the 
integral of \eq{KGL} over $d^4x$ (or equivalently, over $d^3x\, dt)$. 
Because the KG PDE is linear, the units and magnitude of $\Phi$ are immaterial. 

In non-relativistic quantum mechanics, one can establish an equivalence between the  
trajectory-based nonlinear PDE of \eq{nonrelPDE}, describing the solution trajectory
ensemble $x^l(t, {\bf C})$, and the linear TDSE PDE, describing the complex-valued 
wavefunction field, $\Psi(t, {\bf x})$.  It is therefore natural to consider whether or not 
a similar equivalence can be established in the relativistic quantum case---perhaps
leading to the KG PDE, or perhaps to a different wave PDE.   To this end, we seek to convert
the trajectory-based relativistic quantum action of \eq{relaction2} into  a  form 
involving only density fields, rather than explicit $x^\a_{X^\mu}$ quantities.  

Using the constraint  of  \eq{dtaudef}, we first replace the square root quantity in the first 
line of \eq{relaction2} with a factor of unity.  Such a modification is necessary in order to 
remove the explicit  $x^\a_{X^\mu}$ dependences;  however, from the discussion in 
Sec.~\ref{electromagnetism}, 
this change may lead to a fundamentally different Euler-Lagrange PDE.  Next, \eq{rhonat}
is used to replace $\t_\l \,f({\bf C})$ with $\rho(X^\mu)=R(X^\mu)^2$, and Eqs.~(\ref{invdens}),
(\ref{rhostar}),~(\ref{rhogen}), and (\ref{LQalt2}) are applied in order 
to rewrite $L_Q$ in terms of $R(X^\mu)$. The result is:
\ea{
      S  & = &    \int (-mc^2) \!\left [  R(X^\mu)^2  +  {1\over 2} \of{{\hbar \over mc}}^2 
                       \times \right .\nonumber \\
                      & & \left . \sof{\P^i R(X^\mu)} \sof{\P_i R(X^\mu)} \right ] d^3C\, d\l  
                       \label{Sfield3d}
}
As a technicality, it should be noted that the covariant derivatives in \eq{Sfield3d}
refer to the full 4d space, rather than the 3d spatial subspace (as is consistent 
with the fact that $R$ is a 4d scalar density quantity)---despite the fact only the 
spatial components are summed over. 

We now have, in \eq{Sfield3d}, an expression for the action, in terms of 
$R(X^\mu)$ and its spatial ($C^i$) derivatives, which is valid for any set of 
natural coordinates, $X^\mu$.  Note that the time ($\l$) derivatives, $\p_0 R$,
do not appear in \eq{Sfield3d}---as a result of which, the corresponding 
Euler-Lagrange PDE involves spatial derivatives only, and the solution
$R(X^\mu)$ for a given fixed $\l$ value is completely independent from 
that for any other $\l$ value.  The solution $R(X^\mu)$ therefore exhibits 
a constraint in $C^i$ but no constraint in $\l$---exactly the opposite
behavior from that presumed in Sec.~\ref{general-prob}, and which underlies Postulate 1. 
Thus---and despite being linear (in a covariant sense), and expressible 
solely in terms of the $R(X^\mu)$ field as desired---the new PDE is likely
 meaningless, and in any event, is not at all equivalent to \eq{finalPDE}.

A more meaningful result is obtained by introducing a slight modification
to \eq{Sfield3d}---namely, extending the summation to include the time component:
\ea{
      S  & =  &   \int (-mc^2) \! \left [  R(X^\mu)^2  +  {1\over 2} \of{{\hbar \over mc}}^2  
                       \times \right .\nonumber \\       
                       & & \sof{\P^\nu R(X^\mu)} \sof{\P_\nu R(X^\mu)} \left ]
                        \!\of{{1 \over c}} \right .d^4X \label{Sfieldgen}
}
In natural coordinates, \eq{Sfieldgen} now gives rise to an Euler-Lagrange PDE
with both space and time derivatives.  The solution $R(X^\mu)$ field is now
sensibly behaved---at least in most instances. However, even if Postulate 1 is now 
satisfied, Postulate 2 definitely is not; the new $\l$ derivatives 
enter into the determination of the quantum force components, which thus now 
depend on the future states of the system.  Moreover, the PDE is no longer
second order in $\l$, as there are higher order mixed derivative terms that now 
make an appearance.

The primary significance of \eq{Sfieldgen} is that it can be interpreted as a 
completely general covariant expression, with $X^\mu$ an arbitrary set of coordinates.   
Indeed, comparison with \eq{KGL} shows it to be the  {\em generally covariant 
KG action}---apart from an immaterial multiplicative constant.  This
requires that $R$ be identified with $\Phi$ (to within a multiplicative constant,
as discussed), and also that the KG constant $\mu$ take the value
\eb
	\mu = \sqrt{2}\of{{m c \over \hbar}},
\ee
(which is $\sqrt{2}$ times larger than the usual assignment). The corresponding
Euler-Lagrange PDE is the generally covariant version of \eq{KGPDE}; however, 
henceforth, we will find it convenient to work directly in a $X^\mu = x^\a$ inertial 
coordinate frame.

The major lesson learned from this exercise is simply that {\em the KG PDE fails
to satisfy Postulate 2}. In accord with the discussion in Sec.~\ref{classical}, it would appear that
we can therefore dismiss the KG theory, as being unable to provide a viable physical 
description for individual massive particles.  In retrospect, various KG difficulties, 
including those associated with causality,\cite{holland,aharonov69,ranada80} 
have been known in the single-particle  context for many years---despite which, it is still used 
as a basis for many-particle QFT.  In any event, we will discuss some of those drawbacks
here, particularly as they relate to the trajectory formulation. 

With $\rho(x^\a) = \Phi(x^\a)^2$ as presumed above, $\Phi(x^\a)$ must be real  valued,
and so the KG PDE in question refers to electrically neutral particles. This  case 
admits travelling wave solutions of the form
\ea{
	\Phi(x^\a) & = & \cos\sof{{\bf k} \cdot {\bf x} - \omega t}, \label{travel} \\
	{\rm where}\quad \hbar^2 \omega^2  - \hbar^2 c^2 \,{\bf k} \cdot {\bf k} & = & m^2 c^4.
	 \nonumber
}
For the $\Phi(x^\a)$ solutions of \eq{travel}, the wave velocity is well known to be superluminal.  
If probability is conserved as per Postulate 1, and $\rho(x^\a) = \Phi(x^\a)^2$, this in turn
implies quantum trajectories that must also be superluminal---an untenable situation
that should not  be supported in a relativistic theory of massive particles.

On the other hand, KG theory adopts a different definition for the probability density.  
Specifically, the KG flux four-vector is defined as follows:
\eb
	j_\a = {i \hbar \over 2m} \sof{\Phi\, \p_\a \Phi^* - \Phi^*\, \p_\a \Phi}
	\label{KGj}
\ee
In \eq{KGj} above, `$*$' refers to complex conjugation---e.g., $\Phi^* = \Phi$
for the present, neutral particle case.  In general, the KG $j^\a$ satisfies the 
continuity relation, \eq{continuity}. It leads to a scalar probability density,
$\rho(x^\a)^2 = - j^\a j_\a/c^2$, and also a spatial probability density, 
$j^0/c$, both of which are (in general) different than $\Phi^* \Phi$. Indeed,
for neutral particles, the KG $j^\a =0$, in keeping with its interpretation 
as an electrical {\em current} flux.  Thus, for neutral particles, the KG flux adds nothing 
to our understanding of quantum trajectories. 

The KG flux continuity condition becomes non-trivial in the context of 
charged particles, for which $\Phi(x^\a)$ is complex-valued.  In this context,
one would like to associate the trajectory-based $R(x^\a)$ quantity 
with $|\Phi(x^\a)|$, as is done in the non-relativistic case. However,  this 
association does not lead to a KG scalar or spatial probability density that is equal to 
$R^2 = \Phi^* \Phi$, as discussed. On the other hand, the KG forms by themselves
do lead to their own relativistic quantum trajectory ensemble---defined via
$U^\a \propto j^\a$, and propagating under the influence of a quantum potential,
$Q$, that turns out to be proportional to the 4d Laplacian (d'Alembertian) of $R(x^\a)$. 

As sensible as the KG approach to relativistic quantum trajectories---as described 
above---may seem, it is quite problematic.\cite{holland} This is true even if one restricts
consideration to the positive energy solutions only---thus avoiding Problem (1), 
as discussed in Sec.~\ref{intro}. To begin with, the d'Alembertian form of the KG $Q$ clearly 
violates Postulate 2, as discussed.  Second, the KG flux four-vector $j^\a$ (and hence $U^\a$) 
may switch from being timelike to spacelike---thus implying quantum trajectories that
pass through a light cone, which is unphysical. Third, and even more egregious:
the sign of  $j^0$ may change, during the course of the evolution---thus implying 
a trajectory that turns a corner in time, or a particle
that dynamically changes the sign of its charge.  
Finally, the interpretation of the spatial probability density in the two theories---i.e., 
the KG $j^0$ vs. the TDSE $\Psi^* \Psi$---is fundamentally different, with the KG 
$\Phi$ alone insufficient to specify a relativistic quantum state (the time derivative
is also needed). 

It is perhaps remarkable that all of these difficulties  of the KG approach
appear to be remedied though what amounts to a fairly simple fix:  
{\em restrict the derivatives used to 
construct the quantum potential so that these act only on the 3d spatial 
simultaneity submanifolds, rather than the full 4d spacetime}. 
Because, in the relativistic context, the quantum trajectories themselves define 
the simultaneity submanifolds, this restriction also seems to imply that {\em a trajectory-based 
formulation  must  be used in the relativistic context}---or at least, is more fundamental 
than a wavefunction-based  approach.  Though we have not yet succeeded 
in developing a linear relativistic quantum wave equation, it remains
an open question whether or not this is in fact possible.  In addition to the 
reason stated above, another important reason why it may not be possible
has to do with the variation of $\t_\T$ across a simultaneity submanifold, 
and the fact that  relativistic quantum trajectories 
evolve at different rates across the ensemble. Adopting a KG 
quantum trajectory approach, it is exactly this effect that causes 
$j^0/c \ne \Phi^*\Phi$, as well as the other undesirable properties of 
the KG $j^\a$  that have been discussed. 
Alternatively, adopting the present approach, the resultant
relativistic quantum trajectories are evidently well-behaved---but likely 
at the expense of the existence of an equivalent 
linear wave equation. In any case, in the non-relativistic limit---where both 
the wave and trajectory approaches are certainly viable---the variability across
the ensemble of  $\t_\T$, and the associated trajectory evolution rates, disappears.


\section{Examples}
\label{examples}

In this section, we present explicit analytical forms for solutions of 
\eqs{relquantdyn}{finalPDE}---i.e., relativistic quantum 
solution trajectory ensembles, $x^\a(X^\mu)$, expressed in
natural coordinates, $X^\mu=(c\,\T, {\bf C})$---for several  special cases
for which such solutions have been obtained.

\subsection{Quantum inertial motion}
\label{inertial}

The simplest and most obvious example is that of quantum inertial motion. 
In part to justify some of the claims made previously within this document
(e.g., in Secs.~\ref{simultaneity}, \ref{ensemble-time},
\ref{uniformizing}, \ref{conversion}, and \ref{limits}), we here 
work out this example in some  detail. 

We have claimed  that $X^\mu = x^{\a'}$ comprise a quantum inertial solution, where 
$X^\mu$ is a set of uniformizing natural coordinates, and $x^{\a'}$ is any inertial 
coordinate frame, related to $x^\a$ via a Lorentz transformation [\eq{LT}]. Clearly,
$g_{\mu\nu}  = \eta_{\mu\nu}$, $\g^{ij}=\d^{ij}$, and $\g=1$. From the 
second and third lines of \eq{finalPDE}, respectively, we find
$Q=f^\a=0$.  From \eq{tldef}, $\t_\T\!=\!1$, so $\t\!  = \!\T$.

The equation of motion, as expressed by the first line of \eq{finalPDE},
then becomes $x^\a_{\T\T}=U^\a_\T=0$.  It remains only to verify that 
this equation is indeed correct, for our solution ansatz.  From  \eq{LT}, 
\eb
	x^\a_\T = U^\a = c \,{\Lambda^\a}_{0'},
\ee
where ${\Lambda^\a}_{\a'}$ is the inverse of ${\Lambda^{\a'}}_{\a}$.  
The components of the inverse Lorentz transform tensor, ${\Lambda^\a}_{\a'}$,
are constants that do not depend on the spacetime manifold point, $p$.  
Therefore, $U^\a_\T=0$, as required.

\subsection{Exponentially decaying probability density}
\label{exponential}

The case where the probability density decays exponentially is an 
interesting  example---primarily, because it leads to $Q \ne 0$, but $f_i=0$. 
The vanishing quantum force means that  
the quantum trajectories experience unaccelerated
straight-line motion---as in the analogous, non-relativistic case. However
(and quite unlike the non-relativistic case), $Q \ne 0$ implies a time dilation 
or contraction (in this case, contraction) that is also experienced by the relativistic 
quantum trajectories.  This example thus embodies a clean separation of the 
two distinct dynamical effects that arise from the relativistic $Q$, as
discussed in Sec.~\ref{conversion}

Insofar as the quantum trajectory orbits are concerned, these should conform
to parallel straight lines, as in the case of quantum inertial motion.  Without
loss of generality, therefore, we may take those trajectories to be at rest, i.e.,
\eb
	x^l (\T, C^i) = \d^l_i C^i.
	\label{xform}
\ee 
Since $\T$ must be orthogonal to the $C^i$, 
\eq{xform} implies
\eb
	t(\T, C^i) = t (\T). 
	\label{tform}
\ee
Also from \eq{xform}, we can take $\t = t$. 

With the $x^\a(\T, C^i)$ ansatz above, we find:
\ea{
	g_{00} = - \t_\T \quad & ; & \quad g_{0j}=0  \nonumber \\
	g_{i0}=0 \quad  & ; &  \quad g_{ij} = \g_{ij}= \g^{ij} = \d_{ij} 
}
This also leads to $\g \!\!=\!\!1$.
We adopt, for the exponentially decaying density, 
 the explicit form
 \eb
	f({\bf C}) = \exp[-2\, {\bf \kappa} \cdot {\bf C}],
\ee
where $\kappa$ is a constant wavevector quantity, with inverse length units.  
From \eq{Qdef3}, the relativistic quantum potential is then found to be 
\eb
	Q = {-\hbar^2 \over 2m} \,{\bf \kappa} \cdot {\bf \kappa} = {\rm const}
\ee
Note that $Q<0$ everywhere---implying time contraction, rather than dilation,
as indicated above. This behavior is reasonable, considering that exponentially
decaying probability density is associated with quantum tunneling and classically
forbidden regions.

The time contraction effect gives rise, via \eq{tldef}, to a constant value of $\t_\T$ that
is larger than one. This implies that the proper time $\t$ (also equal to $t$) progresses
more quickly than the ensemble proper time, $\T$. In any case, we now have 
an explicit form for \eq{tform}, i.e., 
\eb
	t(\T) = \exp\sof{{1 \over 2} \of{{\hbar \over mc}}^2  {\bf \kappa} \cdot {\bf \kappa} } \T.
	\label{tform2}
\ee

We are now ready to confirm whether \eqs{xform}{tform2} satisfy \eq{relquantdyn}.
The first part of this equation of motion yields
\eb
	{d x^\a \over d\T} = c\, \exp\sof{{1 \over 2} \of{{\hbar \over mc}}^2
	  {\bf \kappa} \cdot {\bf \kappa}} \d^\a_0,
\ee
which is correct, given that $U^\a = (c, 0,0,0)$.  The second part of this equation,
i.e. $U^\a_\T = 0$, is also true, given that $f^\a=0$ because $Q={\rm const}$.

The relativistic quantum trajectory orbits described above are exactly the same
as for the corresponding  non-relativistic application.  Indeed, these trajectory orbits are 
also identical to those for quantum inertial motion---which may now be regarded 
as the limiting case of exponential decay in which the $\kappa^i \ra 0$.  
However,  the new and interesting feature of 
the relativistic trajectory-based theory is the time contraction constant, $\t_\T$.  
In principle, the value of this constant enables one to distinguish between different 
exponential decay applications (including the inertial motion case, 
corresponding to $\t_\T \ra 1$) solely from the trajectory ensemble
itself---i.e., without direct recourse to the probability density function, $f({\bf C})$. 
Such a distinction is impossible in the non-relativistic quantum trajectory theory. 

As an interesting exercise, we consider a sinusoidal, rather than exponential, 
form for $f({\bf C})$---even though such a form does not yield a legitimate solution,
because the probability density $f({\bf C})$ becomes negative. 
This form also gives rise to $f_i=0$ and 
$Q={\rm const}$---but now, $Q$ is positive, rather than negative.  The relativistic quantum
trajectories are still parallel, unaccelerated straight lines, but they experience
(identical) time dilation, rather than contraction.  Interestingly, the resultant
$R(t, {\bf x})$ field is very similar to the KG form of \eq{travel}, except that
the wave velocity is less than the speed of light.

\subsection{Relativistic Gaussian wavepacket}
\label{gaussian}

In the study of the non-relativistic quantum mechanics of free particles,
the time evolution of the 1d Gaussian wavepacket is the canonical example.  
We define this to be a TDSE solution for which the probability density 
has Gaussian form at all times:
\eb
	\rho(t, x) \propto \exp\!\sof{- a(t) [x - x_c(t)]^2}  
\ee
Ideally, an analogous relativistic generalization could be 
derived---giving rise to a nice, well-behaved, nontrivial, benchmark 
analytical solution for the new relativistic quantum PDEs 
of Sec.~\ref{relativistic}. As it happens, though, this is  not possible.

 To begin with,  one is  forced to make a choice, based on 
the following simple question:  what is the most singular 
characteristic of the non-relativistic Gaussian wavepacket?
Specifically, is this the fact that the initial density is of Gaussian form?  
Or alternatively, is it that the initial functional form---whatever that
happens  to be---is preserved over time?  
In the relativistic case, we can satisfy one or the other of these two 
properties, but not both simultaneously.  

In this subsection, we mostly presume that the first consideration
is most important---i.e., that we are working with a relativistic  
wavepacket that is initially of Gaussian form (specifically, a coherent state 
or minimum-uncertainty Gaussian).  
As we shall see, such a wavepacket disperses over time---as may be expected, 
based on the well-known analogous non-relativistic behavior. However, in no sense is
the Gaussian form retained over time. This would appear to render analytical 
solution of the relativistic Gaussian wavepacket  intractible---although a curious
result discussed at the end of this subsection hints that an analytical approach
may indeed be possible. In any event, we shall, in the meanwhile, obtain the solution numerically.
Later, in Sec.~\ref{hyperbolic-wave}, we will then ask the other relevant question, i.e., 
what functional form  {\em is} preserved over time (if any), in relativistic quantum mechanics? It 
 turns out that such solutions do in fact exist, and moreover, that analytical 
 expressions for these can be provided. However, these solutions are singular. 

Before addressing either of the above questions directly, it is worthwhile to consider
exactly {\em why}  it is that the Gaussian form---and only that form---is 
preserved over time in the non-relativistic context. Fortunately, the trajectory ensemble 
approach provides a very straightforward answer. Because of Postulate 1, preservation of the probability density form requires that 
initially uniformly spaced trajectories [i.e., $C = x_0=x(t\!\!=\!\!0,C)$] remain 
uniformly spaced over time.  From a dynamical perspective, this in turn requires 
that the initial velocity and acceleration fields, $\dot x_0 (C)$ and $\ddot x_0(C)$,
be linear in $C$.  The $\dot x_0(C)$ condition is satisfied automatically 
if $\Psi_0(x)=\Psi(t\!=\!0,x)$ is itself a Gaussian [as opposed to just $\rho_0(x)$]. 
In the overwhelming majority of applications, the initial $\Psi_0(x)$ is taken to 
be a coherent state Gaussian, for which $\dot x_0(C) = {\rm const}$ is 
independent of $C$ (i.e., all trajectories are instantaneously moving in unison).  
The $\ddot x_0(C)$ condition is satisfied if and only if the initial force field 
is linear in $C$.  For a free particle, only quantum forces are present, and so a 
linear force field requires that $Q_0(C) = Q(t\!\!=\!0,C)$ exhibit a quadratic 
dependence on $C$. The only  $\rho_0(x)$ functional form that can 
give rise to such a $Q_0(C)$ is the Gaussian.

From \eq{Rdef}, an initially Gaussian $\rho_0(x)$ implies a Gaussian
$f(C)$, because $\g = 1$ at $t=0$ (since $C = x_0$ is presumed, and 
$\g=\g_{11}$ because the application is 1d).  Of
course, $f(C)$ is preserved over time---but this property is not  unique to
the Gaussian alone; it is always true, due to Postulate 1. 
Rather, the fact that both sides of \eq{Rdef} maintain a Gaussian form for
all $t$ implies something special about $\g$---namely, that 
$\g(t,C)=\g(t)$ is independent of $C$.  The value of $\g(t)$ indicates 
the extent of Gaussian dispersion or broadening present at time $t$. 
Thus, if the initial Gaussian is taken to be a coherent state, 
$\g(t) \ge 1$ for all $t$, with the equality holding only at $t=0$. 

Let us now apply a similar analysis to the relativistic free particle problem.  
Without substantive loss of generality, we presume an initially {\em stationary} 
coherent state Gaussian wavepacket of the following form:  
\ea{
		f(C) & \propto & \exp[-a C^2],  \label{Gaussian} \\
		{\rm with} \quad  t_0(C)= t(\T\!\!=\!0,C)&  = &  0    \nonumber \\ 
		                                  x_0(C)  =  x( \T\!\!=\!0,C) &  = &  C  \nonumber \\
		 t_{\T,0}(C)  = \left . \prt t {\T} \right |_{\T\!\!=\!0}  &  =  & 
		                                          \exp\sof{-{Q_0(C) \over mc^2} }   \nonumber \\
		                              x_{\T,0}(C)  =  \left . \prt x {\T} \right |_{\T\!\!=\!0}  & = & 0 \nonumber   \\        
		{\rm and} \quad  Q_0(C)  = Q(\T\!\!=\!0,C) &=& -{\hbar^2 \over 2m} (a^2 \, C^2 -a)
		                      \nonumber             	                                    
}
Again, the fact that $f(C)$ {\em per se} remains Gaussian over time is not special,
as this property would be true for any relativistic quantum wavepacket,  by virtue of 
Postulate 2.   In what sense, then, should the question of the preservation of the
Gaussian form be addressed?

To this end, it is natural to consider the usual spatial probability density,
$j^0(t,x)/c$. However, for fixed $t>0$, this function cannot be Gaussian. 
The reason is that the linear  (with respect to $x$) velocity field, $\dot x(t,x)$, 
that this would entail would require the exterior trajectories in the ensemble to 
travel superluminally. The one exception would be the case of parallel
trajectories; but in this case, the probability density would be uniform or exponential, 
rather than Gaussian (see Secs.~\ref{inertial} and~\ref{exponential}).  
 Alternatively, it is more natural in the relativistic context to adopt the 
 simultaneity submanifolds---i.e., the contours of 
constant $\T$ rather than $t$---as the subspaces over which the Gaussian form 
might be preserved.  The relevant probability density quantity is thus $\rho^*(X^\mu)$.  
The {\em a priori} reason why this approach might have a chance of being
successful is because the simultaneity submanifolds fan outwards away from the 
$x$ axis, so that the interior trajectories are evaluated  at earlier $t$ relative to 
the the exterior trajectories (see Fig.~\ref{fig-Gaussian}).   The latter are thus afforded more time 
to spread themselves out. 

The condition that the trajectories be distributed uniformly across a given 
simultaneity submanifold, labelled by $\T$, is that $\g(\T,C)=\g(\T)$ be independent of 
$C$---in exact analogy with the non-relativistic case.   In order that this 
condition be preserved over time, one must have, in addition to the linear 
initial conditions of \eq{Gaussian}, a linear acceleration field, where the acceleration
is taken with respect to $\T$---i.e., 
\eb
	\prtsq {x^\a}{\T} = A^\a + B^\a \,C, 
	\label{accelcond}
\ee
where the vectors $A^\a$ and $B^\a$ are $C$-independent constants. 
In contrast, the Gaussian form of \eq{Gaussian} provides a linear initial 
acceleration field, but with respect to the {\em proper time} $\t$, rather than 
the ensemble time $\T$.  The two time derivatives are related to each 
other by a factor of $\t_\T$, which depends on $Q$ via \eq{tldef}, and 
thus on $C$ in nonlinear fashion.   The initial Gaussian wavepacket 
therefore does not satisfy \eq{accelcond}, and so the initially  
Gaussian form of  $\rho^*(X^\mu)$ is not preserved over time. 

Once again, we see in this example a manifestation of the principal difference 
between the non-relativistic and relativistic quantum  theories---i.e., the 
distinction that exists in the latter case between $\t$ and $\T$, both of 
which are dynamically relevant.  As discussed in Sec.~\ref{conversion} 
and~\ref{exponential}, this 
essentially corresponds to the distinct dynamical roles played 
by the quantum force, $f^\a$, and the quantum potential, $Q$.  This distinction
can give rise to some quite interesting and competing dynamical effects,
even for the simple case of the wavepacket dynamics corresponding to
the initially Gaussian form of \eq{Gaussian}. For simplicity, we continue to
refer to this example as the ``relativistic Gaussian wavepacket,'' even though
it is understood that the Gaussian form is not preserved over time. As stated previously,
the relativistic Gaussian wavepacket propagation can be computed numerically.
Accordingly, we now describe the procedure that we have  used to do so. 

Since the problem is ``1d'' in the sense that there is only  one spatial 
coordinate, the $\tilde \g$ spatial metric ``tensor'' is 
really just a  single number---i.e., $\tilde \g = \g = \g_{11}$.  Using this fact, together
with Eqs.~(\ref{gdef}), (\ref{Qdef3}), (\ref{tldef}), (\ref{finalPDE}), and 
(\ref{Gaussian}), one can express the dynamical PDEs explicitly in
terms of the $\T$ and $C$ partial derivatives of the $x^\a = (c\, t, x)$,
as follows:
\ea{
      -{1 \over m}\, {t_C \over \g}\, \prt{Q}{C}  & = &
      e^{Q \over mc^2}\, \prt{}{\T}\! \sof{e^{Q \over mc^2} \,t_\T}   \label{Gaussdyn} \\
     -{1 \over m} \,{x_C \over \g}\, \prt{Q}{C} & = &
     e^{Q \over mc^2}\prt{}{\T} \! \sof{e^{Q \over mc^2} \, x_\T} ,   \nonumber \\
    {\rm where} \quad Q & = &  {-\hbar^2 \over 2m}\sof{{e^{{a C^2 \over 2}} \over \g^{1/4}}} \prt{}{C}
    \! \sof{ {1 \over \g^{1/2}} \,\prt{}{C}\!\sof{{e^{-{a C^2 \over 2}} \over \g^{1/4}}}  }\nonumber \\
     {\rm and} \quad \g & = & x_C^2 - c^2 \,t_C^2 \nonumber
}

We have applied \eq{Gaussdyn} to a specific Gaussian wavepacket
that has already been considered previously, in the non-relativistic quantum 
trajectory context.\cite{poirier10nowave}  For this example, the relevant
parameters are as follows: $m=\hbar=1$; $a=1/2$.  We have also 
chosen coordinate ranges that are essentially identical to those of the previous
non-relativistic calculation---i.e., $0 \le \T \le 10$, and $-5 \le C \le 5$.
The $C$ range encompasses almost all of the total probability of the 
ensemble, whereas the $\T$ range is sufficiently extensive to incorporate
a very substantial amount of wavepacket broadening. As for the speed of light, 
$c$, this is here taken to be an adjustable parameter, so as to enable 
consideration of a range of behaviors, from the non-relativistic limit 
($c\!=\!100$) to the ultra-relativistic limit ($c\!=\!1$).  We will 
mainly focus on the $c=3$ case, as a specific example that manifests 
very substantial relativistic behavior.  

The PDEs of \eq{Gaussdyn} were solved using Mathematica's NDSolve 
routine, using the method of lines with 25 spatial grid points, but without any 
explicit specification of boundary conditions.    For $c\!=\!3$,  the entire calculation 
took 2.14 s on a 2.7 GHz Intel Core i7 CPU.  Of course, we have no 
analytical solution for comparison; however, the numerical solution so obtained
satisfies the PDEs for both $c\, t$ and $x$  to an absolute accuracy of 
1.4 $10^{-6}$ or better, throughout the coordinate ranges indicated above.
For the $c\!=\!100$ calculation, the computed quantum trajectories are virtually 
indistinguishable from those of the previous non-relativistic calculation, and
the simultaneity submanifolds are likewise nearly perfectly horizontal (constant-$t$)
lines.  The relativistic solution therefore does indeed reduce to the non-relativistic
solution in this limit, as predicted in Sec.~\ref{limits}.  For the ultra-relativistic case, $c\!=\!1$,
the relativistic quantum trajectories very quickly begin to fan out, after which 
they also approach their terminal velocities quickly.  In certain respects, this limiting
behavior is the same as that of an infinitely narrow relativistic Gaussian, as discussed 
in Sec.~\ref{limits}.  Numerical solutions for various other $c$ values in the $1 \le c \le 100$ 
range have also been computed, but these will not be reported on here.

The $c\!=\!3$ relativistic quantum trajectories, and associated simultaneity 
submanifolds, are those presented in Fig.~\ref{fig-Gaussian}.  This example exhibits very 
marked relativistic effects---as manifest, e.g.,  in the curvature of the simultaneity  
submanifolds, as well as in the extensive distortion present in the upper corners 
of the figure, where the  trajectories  are starting to approach the speed of light.  
For this example,  the trajectories remain vertical and parallel for a little while at the 
start of the propagation, before fanning outward from each other. On the other 
hand, the simultaneity submanifolds begin curving outward immediately, 
as soon as $\T\!>\!0$. This is a clear manifestation of the two distinct dynamical 
effects, as discussed above, and in Sec.~\ref{conversion}.  Specifically, the curvature of
the simultaneously submanifolds is due to the $C$ variation of $\t_\T$---which, 
due to the latter's dependence on $Q$, is in 
play even at $\T\!=\!0$. In contrast, the fanning out of the trajectories is due to 
$f^\a_Q$, which influences the trajectory dynamics via the second, rather than
the first, derivative in $\T$. [Note that we have returned to using a $Q$ subscript
when referring to the quantum force, to avoid confusion with the spatial scalar
probability density, $f({\bf C})$.]

\begin{figure}
\includegraphics[scale=0.60]{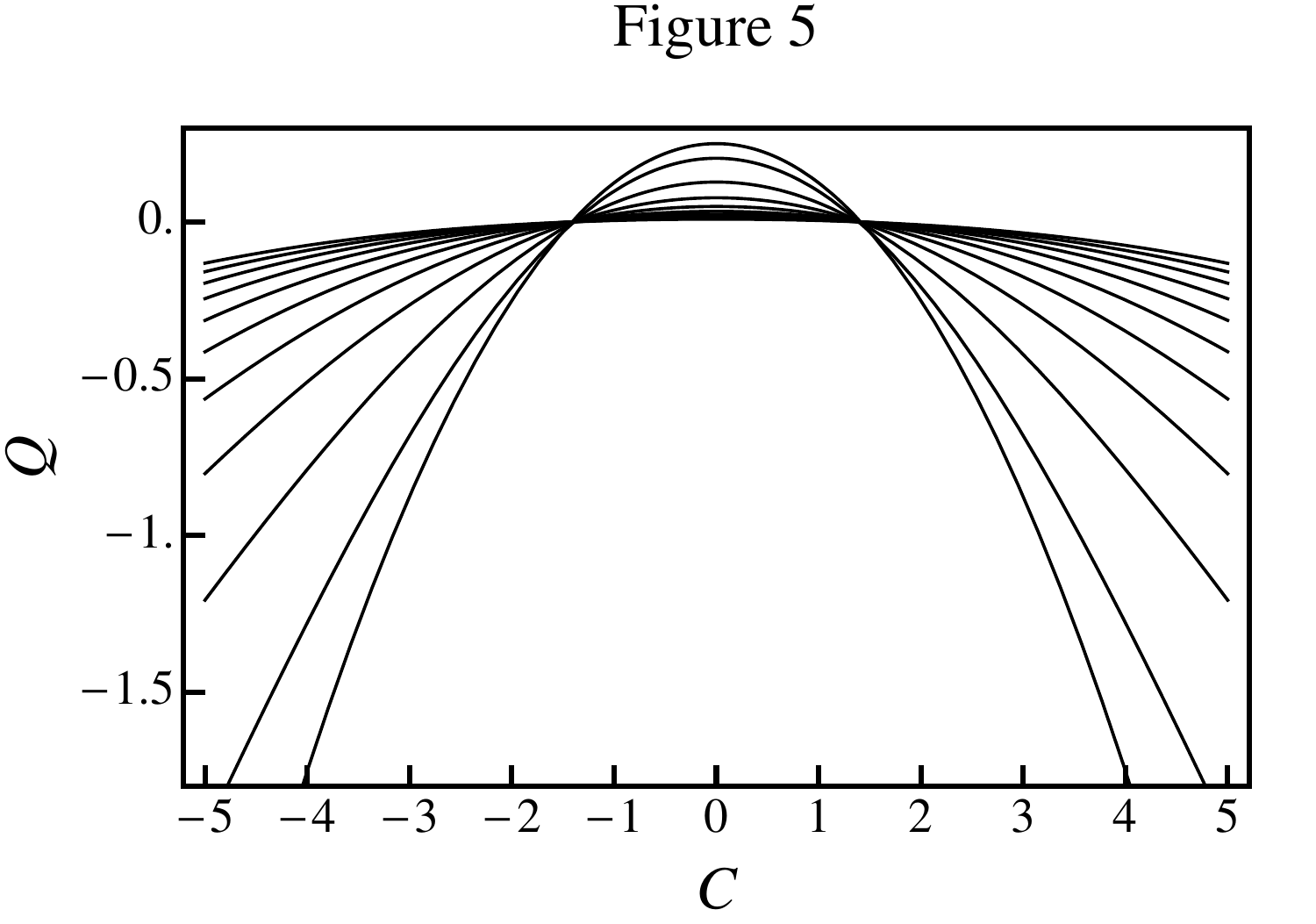}
        \caption{Quantum potential, $Q$, for the 1d  $c\!=\!3$ relativistic Gaussian wavepacket of 
Sec.~\ref{gaussian}, 
as represented in the natural coordinate frame, $X^\mu=(c \, \T , C)$.  Each curve represents 
$Q$ as a function of $C$ for fixed $\T=\{0,1,2 \ldots,10\}$, with increasing $\T$ values corresponding to 
flatter curves. $Q\!>\!0$ and $Q\!<\!0$ represent, respectively, the classically allowed and forbidden
regions of spacetime. These are demarcated by two special ``reference'' trajectories, 
$C_{{\rm ref}} = \pm \sqrt{2} $, along which $Q\!=\!0$ for all time.}
        \label{fig-Q}
\end{figure}

The center-most, $C\!=\!0$ trajectory in Fig.~\ref{fig-Gaussian}---which is the one with the 
greatest probability density---also represents a striking demonstration
of the dynamical distinction described above.  Specifically, this trajectory
experiences no quantum forces---and therefore no ``acceleration,'' in the usual
sense involving $\t$.  As a consequence, this trajectory is straight, rather than curved
like all of the others in the ensemble. On the other hand, this trajectory does 
experience pronounced  time dilation---more so than any other trajectory, in fact, 
because the value of $Q$ is largest there.  Moreover, there is a marked variation 
in the magnitude of the time dilation experienced, over the course of this trajectory's 
evolution.  This variation is associated  with a corresponding variation in 
$Q(\T,C\!\!=\!0)$, that comes about due to wavepacket broadening (see the 
discussion below pertaining to Fig.~\ref{fig-Q}). 

\begin{figure}
\includegraphics[scale=0.60]{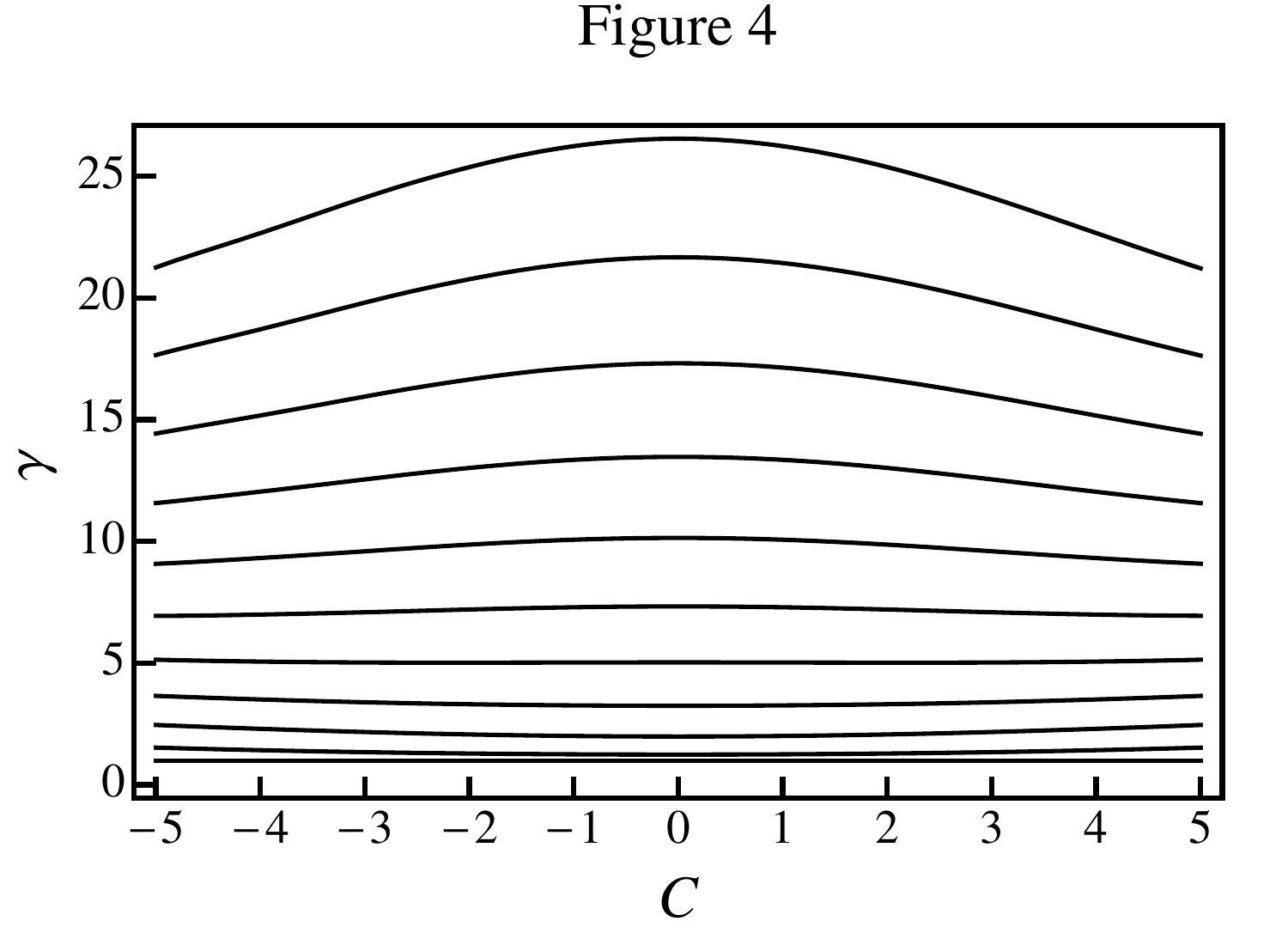}
        \caption{Spatial metric tensor, $\g=\g_{11}=g_{11}$, for the 1d  $c\!\!=\!\!3$ relativistic
Gaussian wavepacket of Sec.~\ref{gaussian}, as represented in the natural coordinate frame, 
$X^\mu=(c \, \T , C)$.  Each curve represents $\g$ as a function of $C$
for fixed $\T=\{0,1,2 \ldots,10\}$, with increasing $\T$ values corresponding to 
increasing $\g$. At early times, $\T\!>\!0$, the curves bow upward, but at later times
they bow downward.}
        \label{fig-gamma}
\end{figure}

Further dynamical insight into the relativistic Gaussian wavepacket dynamics
may be gained from Fig.~\ref{fig-gamma}. This is a plot of $\g$
as a function of $C$, for various fixed values of $\T$, for the $c\!=\!3$ 
example.   The primary feature of this figure is that
the $\g(C)$ curves increase in magnitude with increasing $\T$---thus 
indicating the expected wavepacket broadening, with respect to the 
simultaneity submanifolds. Note, however, that this increase is not uniform across 
$C$---i.e.,  the curves are not horizontal lines.  If they were, this 
would imply preservation of the Gaussian form over time, which as 
discussed, does not occur (except in the non-relativistic limit).  
Conversely, the curvature or bowing of these curves---which for this example, 
is seen to be rather pronounced---is an indication of relativistic quantum 
dynamical effects.

Regarding the bowing of the $\g(C)$ curves, a curious feature may be observed 
in Fig.~\ref{fig-gamma}, which is that the direction changes over time.
The early curves bow upward, whereas at later times, the 
curves bow downward.  This can be explained, yet again, as a competition
between between $Q$ and $f^\a_Q$ dynamical effects.  At early times,
all trajectories are moving (really at rest) in unison; their acceleration has not
yet had a chance to manifest as a fanning out of velocities.  However, even at
$\T\!\!=\!0$, the local proper time evolves at a higher rate 
towards the exterior fringes of the ensemble (i.e., towards larger $|C|$).   
The early simultaneity submanifolds 
thus curve away from the $x$ axis, resulting in larger intervals between
nearby trajectories that that lie further from $C\!\!=\!0$---and thus, in upward-bowing
$\g(C)$ curves.  Over time, acceleration gives rise to trajectory fanout, as discussed.  
However, as the velocities of the exterior  trajectories reach the order of the speed 
of light, further broadening of the wavepacket  is relativistically hindered, and so the
resultant $\g(C)$ curves bow downward, rather than upward.   

The dynamical behavior of the quantum potential, $Q(\T,C)$, is also worthy of
discussion. In Fig.~\ref{fig-Q} are presented curves of $Q$ as a function
of $C$, for various fixed values of $\T$ (the same values as in Fig.~\ref{fig-gamma}), for
the $c\!=\!3$ relativistic Gaussian wavepacket. At $\T\!\!=\!0$, $Q\!\!=\!Q_0(C)$ 
exhibits the usual negative quadratic pattern [\eq{Gaussian}], well known 
in the context of non-relativistic Gaussian wavepackets.  The $Q_0(C)$ curve is 
positive in the interior of the ensemble and negative in the exterior, with the 
turnover, $Q_0(C_{\rm ref})=0$, occurring at 
$C_{\rm ref}=\pm \sqrt{1/a}= \pm \sqrt{2}$. These points thus serve as the 
instantaneous demarcation between ``classically allowed''  and 
``classically forbidden'' regions. A remarkable feature to emerge from  Fig.~\ref{fig-Q} 
is that {\em these demarcation points are the same for all values of $\T$}.  
In other words, the boundary between allowed and forbidden regions is 
marked by two special trajectories, $C_{\rm ref} =\pm\sqrt{2}$. 
We refer to these as ``reference trajectories,''  because of the
fact that $Q\!=\!0$ everywhere along these trajectories, implying from \eq{tldef}
that $d\t = d\T$.  The proper time, $\t$, as measured along a reference trajectory,
may therefore be used as the dynamical ensemble proper time, $\T$, for
the entire trajectory ensemble.  

It should be emphasized that the existence of reference trajectories  for 
a given trajectory ensemble is not to be generally expected. We have no 
general existence proof---nor, indeed, has their existence been mathematically 
proven even for the relativistic Gaussian wavepacket (although for the $c\!=\!3$ case
explicitly considered here, $Q(\T,\pm\sqrt{2})=0$ has been established numerically 
across all $\T$, i.e. not just for the eleven $\T$ values indicated in Fig.~\ref{fig-Q}).  That 
reference trajectories do seem to exist in the relativistic Gaussian case is therefore quite
special---perhaps indicating that in some sense, the Gaussian form is
preserved, after all. This important property
may also provide insight that ultimately makes it possible to arrive at an analytical  
solution, although to date such efforts have  not borne fruit.   

Figure~\ref{fig-Q} also provides---via \eq{tldef}---information pertaining to the local time dilation
and contraction across the ensemble.  We see that these effects are greatest at the initial time, 
$\T\!\!=\!0$---perhaps counterintuitively, given that the simultaneity submanifold
(i.e., the $x$ axis) is least distorted there.  Over the course of time, wavepacket 
broadenening leads to a reduction in the magnitude of $Q(C)$ across all of the
trajectories, $C$, and thereby, to a reduction in the time dilation/contraction effect.  
As discussed, the trajectories in the classically allowed region between the reference trajectories
experience time dilation (with the $C\!=\!0$ trajectory experiencing this to the
greatest extent), whereas trajectories in the forbidden region experience time contraction.

\subsection{Hyperbolic wavepacket solutions}
\label{hyperbolic-wave}

In Sec.~\ref{gaussian}, we have seen that the relativistic ``Gaussian'' wavepacket does not
preserve its Gaussian form over time, in the sense that $\rho^*(\T, C)$ is not
Gaussian and $\g(\T,C)$ depends on $C$, for fixed $\T\!\ne\! 0$.  This situation 
begs the question: is there {\em some other} choice of  $f(C)$, for
which the analogous properties do hold?  We show in this subsection that such
solutions do indeed exist, and moreover, that they can be obtained analytically. It should
be stated at the outset, however, that these solutions contain singularities. The solution
ensembles also contain trajectories that move along a light cone. Since every 
simultaneity submanifold includes at least one such trajectory moving 
at the speed of light, any choice for the initial  $\T$ value will
lead to initial conditions that necessarily violate the requirements 
for a viable solution, as laid out in Sec.~\ref{basic}.  This situation by itself
provides sufficient {\em a priori} cause to discard these ensembles; hence
in that sense, it does not matter that these solutions develop singularities
at later or earlier times, because they do not technically belong to our theory.  
Nevertheless, these solutions are still worth considering because
they are simple and analytical, and because they serve as asymptotic limits
for other solution ensembles that are in fact  viable. For this reason, it will 
prove instructive to work through them in some detail. 

 As per the discussion in Sec.~\ref{gaussian}, our starting point will be the condition that 
 $\g(\T,C) = \g(\T)$ be independent of $C$.  Actually, it will prove beneficial to first 
 consider the special case where 
\eb
          \g(\T) = {\rm const} = 1\qquad {\rm for}\,\, {\rm all}\,\,\T. \label{gone}
\ee         
 This corresponds to 
 no wavepacket broadening at all, in the sense of the simultaneity submanifolds.  
 However, such solutions---should they prove to exist---will still manifest 
 broadening in $(t,x)$ space, due to the increasing curvature of the simultaneity
 submanifolds with increasing $\T$ (Sec.~\ref{gaussian}).  In any case, the
 fact that $\g\! =\! 1$ for all $\T$ and $C$, together with \eq{Qdef3},  implies that 
 the quantum potential exhibits no $\T$ dependence---i.e., $Q(\T,C)= Q_0(C)$
 for all $\T$.  The same must be true for 
 $f^{\mu=1}_Q = f^Q_{\mu=1} = f_Q= -\p Q/\p C=-Q_C$.      
 
 The fact that the force, $f_Q(C)$, depends only on $C$ implies that 
 {\em all trajectories in the solution ensemble undergo constant acceleration}, 
 in the usual proper time sense. In standard relativity theory, the general form for a 
 trajectory undergoing constant acceleration is well known;\cite{rindler,dolby03} 
 such trajectories trace out hyperbolic orbits.  Adopting the initial conditions from 
 the second and third lines of \eq{Gaussian}, the solution ensemble must thus be of the 
 following form:  
 \ea{
 	t(\t,C) &  =  & {m c \over f_Q(C)} \,\sinh\! \sof{{f_Q(C) \,\t \over mc}} \label{hyperbolic} \\
 	x(\t,C) &  =  & {m c \over f_Q(C)} \of{\cosh \! \sof{{f_Q(C)\, \t \over mc}}-1} + C   \nonumber 
 }
Next, we convert proper time to ensemble time, using \eq{tldef} and the fact that 
$t=\t=\T=0$. The result is
 \ea{
 	t(\T,C) &  =  & {m c \over Q_C} \,\sinh\! \sof{{Q_C \over mc}\,\exp\!\of{{-Q\over mc^2}} \T},
	      \label{hyperbolic2} \\
 	x(\T,C) &  =  & -{m c \over Q_C} \of{\cosh \!\sof{{Q_C \over mc}\,\exp\!\of{{-Q\over mc^2}} \T} -1}
	        + C.   \nonumber 
 }

It remains only to find a $Q(C)$ such that \eqs{gblock}{gone} are
both satisfied. It can be shown that any such $Q(C)$ satisfies 
\eb
	f_Q(C)^2 = - m c^2 f_Q'(C). 
\ee
The trivial solution, $f_Q(C)=0$, gives rise to the uniform and exponential
cases considered in Secs.~\ref{inertial} and~\ref{exponential}, respectively. 
 Otherwise, the most general solution is 
\eb
	f_Q(C) = {m c^2 \over C}, \label{fsol}
\ee
apart from an immaterial shift in $C$. This leads to 
\eb
	Q(C) = -mc^2 \log[B\,C] \quad ; \quad \t_\T(C) = B\, C,
	\label{Qsol}
\ee
with $B$ an arbitrary constant. Note that $Q(C)$ is singular 
at the origin. 

Finally, substitution  of \eqs{fsol}{Qsol} into \eq{hyperbolic2} 
leads to remarkably simple explicit forms for the resultant $\g\!=\!1$ solutions:
\ea{
 	t(\T,C) &  =  & {1\over c} \,C \sinh\! \sof{c\, B\,\T}
	      \label{hyperbolic3} \\
 	x(\T,C) &  =  & C \cosh \!\sof{c\, B\, \T}   \nonumber 
 }
The simultaneity submanifolds associated with these solutions are 
{\em straight lines}---all intersecting at the origin, which is therefore a
singular point.  This situation, and \eq{hyperbolic3} itself, are familiar, in the context of
constructing comoving frames for relativistic particles undergoing
constant acceleration.  As discussed in Sec.~\ref{intro}, this leads to multiple
reoccurences of the same spacetime events (the origin in this case),
and to simultaneous forward ($C>0$) and backward ($C<0$) time
propagation. 

Of course, it is exactly this type of situation that we sought to {\em avoid} 
with the present formulation.  Again, though, we emphasize that the \eq{hyperbolic3}
solution is strictly speaking not a part of our theory, as it  can be dismissed based on
initial condition considerations alone. More specifically, for the $\T=0$ simultaneity
submanifold,  the initial conditions are
not well-defined at the origin. In effect, there are two trajectories 
emanating from this point, forming a light cone, and thus 
moving at the speed of light.  In any case, $\T\!\!=\!0$ is ruled out 
as a viable initial submanifold.  At other $\T$ values, the situation is
even worse, because  \eq{hyperbolic3} is only defined outside of the light 
cone---thus leaving gaps where there is no solution at all.  
As a result, any $\T \ne 0$ also fails to serve as a legitimate initial submanifold,
in the sense of Sec.~\ref{basic}.

Having obtained a solution to the $\g\!=\!1$ problem---albeit a singular 
one---we next consider the case where $\g\!=\!\g(\T)$. One
obvious solution of this type can be easily obtained from  \eq{hyperbolic3},
upon realizing that
$g_{00}\!=\!g_{00}(C)\!=\!-B^2 C^2$ is independent of $\T$.  This is relevant when
one considers that simultaneous exchange of $(c\,t, x)$ and $(c \,\T, C)$ will
flip the roles of $g_{00}$ and $g_{11}$---thus leading to a new dynamical 
solution undergoing constant (and in fact, zero) acceleration, but for which
$g_{11}=\g=\g(\T)$.  The analytical form of the new solution is:
\ea{
 	t(\T,C) &  =  &  \T  \!\cosh\! \sof{A \,C}
	      \label{hyperbolic4} \\
 	x(\T,C) &  =  & c \,\T\! \sinh \!\sof{A\, C}   \nonumber 
 }
 
Now it is the  trajectories of \eq{hyperbolic4} that are the straight lines 
intersecting at the origin, with the simultaneity submanifolds 
the associated hyperbolae. 
Also, it is now the region {\em inside} rather than outside
the light cone, that is accessible to the system. In any case, this
solution has the same sorts of difficulties as \eq{hyperbolic3}, and 
can be similarly dismissed. Since the two solutions,  
\eqs{hyperbolic3}{hyperbolic4} explore complementary 
regions of space, one might consider combining them 
 together.  Such an approach would
``almost'' work, from the perspective of providing reasonable 
results throughout most of spacetime.  However, the origin, 
and the light cone itself, would still be problematic---thus again
making it impossible to find a suitable set of initial conditions
for any choice of $\T$.

As we have seen, the solution ensembles developed in this section
are not globally viable. Locally, however, they may still serve a 
useful purpose,  particularly as asymptotically limiting forms. 
For example, a relativistic Gaussian wavepacket in the 
ultrarelativistic limit of small $c$ or large $a$, is characterized
by trajectories that are observed to originate from a small
region in space, and to approach their asymptotic 
straight line forms very quickly.  The 
extreme limit of this behavior would thus correspond to  an initially 
``Dirac delta function'' wavepacket, i.e. to 
\eq{hyperbolic4}.


\section{SUMMARY AND CONCLUSIONS}
\label{conclusion}

In a series of previous articles, the trajectory-based formulation
of non-relativistic quantum mechanics was developed in some
detail.\cite{bouda03,holland05,poirier10nowave,holland10,poirier11nowaveCCP6,poirier12nowaveJCP,poirier12ODE} What that work offers, in effect, is an 
alternative---to the usual non-relativistic theory based on 
the wavefunction, $\Psi$, and the TDSE PDE governing its time
evolution.  Of course, there are other alternate formulations---and
interpretations---of non-relativistic quantum theory,\cite{styer02,madelung26,bohm52a,bohm52b,holland,wyatt,durr92,berndl95,einstein35,ballentine70,home92,everett57,dewitt70,wheeler} some of which have been around for a very long time.  
Almost all of these make use of $\Psi$, or of quantities derived 
directly from it.  The trajectory-based approach stands somewhat 
apart from these others, because it is formulated independently
of $\Psi$ and of the TDSE.   Instead, the starting point  is 
the path ensemble ${\bf x}(t, {\bf C})$, from which the solution ensemble
is singled out (via the standard Euler-Lagrange prescription) as that
which extremizes the non-relativistic quantum action. This leads to 
the trajectory-based non-relativistic quantum PDE, which happens to be
formally equivalent to the TDSE.  Thus, even though the  resultant
quantum trajectories turn out to be identical to those of Bohmian 
mechanics, the theoretical formulation is completely different---as
Bohm's approach still requires  an external $\Psi$ field, 
to serve as the ``pilot wave''  governing the quantum trajectory dynamics.

As intriguing as a wavefunction-free formulation of non-relativistic
quantum mechanics may be,  the fact that the trajectory-based
PDE is mathematically equivalent to the TDSE---and to all
other non-relativistic quantum formulations, for that matter---suggests
that there may be no empirical means available that can distinguish 
one such theory from another.  They may be all equally ``correct,'' insofar 
as experiment is concerned.  This statement is not meant to diminish
 the value of the various  competing  approaches;   
 each has proven  to have its own peculiar advantages and 
 disadvantages, vis-\`a-vis the  interpretation, computation, 
 and formulation of the non-relativistic quantum theory.  

On the other hand, the approach presented in this document represents a very 
significant departure from all of these previous efforts.  To the author's best 
knowledge, the present work represents the {\em only} (evidently) viable 
relativistic quantum formulation for a single spin-zero particle with 
mass.  In this context, the Klein-Gordon (KG) theory is the first and 
best known competing formalism, although its  problems as a single-particle 
theory are profound (Sec.~\ref{KG}, and citations therein).  
More to the present point, however, {\em the KG 
theory and the present theory make different empirical predictions};
they are {\em not} mathematically equivalent, and so it should be possible
to distinguish between them experimentally.  Therefore, regardless of
numerical convenience or interpretational elegance, the issue of
which theory is ``right'' is one that can in principle be settled in the laboratory.

No experiment will confirm all of the predictions of KG theory---since
these include certain unphysical phenomena such as scattering
probabilities that are greater than one.\cite{bohm,holland,aharonov69}  
Likewise, the validity of relativity is beyond reproach, and so no experiment
performed in the relativistic limit will confirm the predictions of non-relativistic
quantum theory---regardless of the particular formulation that is adopted.  
Such experiments may, however,  be able to validate the present 
relativistic quantum theory, or something similar to it.  If this were to happen, then 
apart from the obvious direct value,  it could also shed light on matters of quantum 
interpretation and ontology that might otherwise
linger in the realm of metaphysics rather than science. 

This prospect bears some further discussion. Working from the assumption 
of experimental validation of the sort posited above, one must consider 
the possibility that there might exist {\em other} theoretical formulations, not yet 
devised, that would lead to formally identical predictions as the present 
theory---as indeed, we have seen to be the case in the non-relativistic limit.  
For example, one might well attempt to relativize the other non-relativistic formulations, 
in analogy with what we have done here for the trajectory-based approach. Such 
a strategy is likely to fail, however---because the other formulations 
rely explicitly or implicitly on the TDSE for $\Psi$, whose relativistic generalization
is presumably the KG equation.  Although the prospect of a viable single-particle
relativistic linear wave equation is not entirely ruled out, it seems 
quite unlikely (Sec.~\ref{KG}, and discussion below);  in any case, no such equation has 
materialized after more than eighty years. Without such an equation, however, 
it is not clear how  the other, $\Psi$-based non-relativistic formulations are to be
relativized, as the linear superposition postulate is not satisfied.\cite{bohm,messiah}

Under such hypothetical circumstances, the present formulation---and any 
associated  ontological interpretation---would become the {\em only} 
viable single-particle theory able to explain experimental observation, as 
opposed to one amongst several equivalent competing theories.  
As discussed in Sec.~\ref{simultaneity}, the 
present approach indeed suggests its own natural ``many worlds''  
interpretation---wherein each trajectory in the ensemble constitutes a distinct
world, or universe.  This is very different from  ``conventional'' 
many worlds theory,\cite{everett57,dewitt70,wheeler} however. For one thing, 
the trajectory worlds {\em do} interact with 
each other (which is, indeed, the source of all quantum behavior); for another,  
the worlds do not branch over time (new trajectories are not ``born,'' e.g., as a result
of measurement, but exist for all time).  A further discussion would be out of place 
in this document, but can be found in our previous 
work,\cite{poirier10nowave,poirier11nowaveCCP6,poirier12nowaveJCP}
 and in future work specifically addressing the quantum measurement ramifications 
 of the trajectory-based approach. 

After the above, admittedly speculative discussion, it now behooves us to provide 
a summary of what has actually been achieved here thus far.  The key ideas, in 
rough order of logical dependence, are listed below:
\begin{itemize}
\item General covariance
\item Trajectory ensembles
\item Simultaneity submanifolds
\item Postulate 1: probability conservation along trajectories
\item Postulate 2: dynamical forces do not depend on future system states
\item Universality of the quantum potential
\item Principle of least (or extremal) action
\end{itemize}
These are the most essential elements, leading to a nearly unambiguous 
theory of relativistic quantum mechanics, as presented in this work. 

Of course, the Einstein Equivalence Principle,\cite{carroll,weinberg} 
 or more broadly, the principle of
general covariance, pervades everything that we do here---restricting possibilities, 
and otherwise guiding towards the relativistically proper forms for all quantities. 
The idea of ensembles of trajectories that foliate spacetime is borrowed from
the non-relativistic  theory, but is equally natural in the relativistic context,
where it is arguably even more important.  The reason is that it leads directly to the
construction of  simultaneity submanifolds for  accelerating particles---a 
central idea of this work, crucial to all of the subsequent development presented here.

A rigorous, global definition of
simultaneity is problematic for classical relativistic theory, because a single
trajectory can explore only a small portion of spacetime. It is not clear that any such
viable notion existed previously---despite certain clever prior developments
in this direction, involving intersecting light cones emanating from the worldline,
and the like.\cite{dolby03}   In a trajectory {\em ensemble} approach, however, 
the construction of simultaneity submanifolds is extremely natural and 
unambiguous. In this respect, the quantum mechanical generalization of
relativistic theory actually makes things much easier. 

Postulates 1 and 2 provide important insight  into the kinematics
and dynamics associated with the trajectory ensemble time evolution. One 
of the most surprising conclusions (at least for the author) to come out of this 
work is the idea that the KG equation does not satisfy Postulate 2---as 
reasonable as both the equation and the postulate might seem.  The latter 
is, after all, satisfied by relativistic and non-relativistic classical mechanics, 
as well as by non-relativistic quantum mechanics; it really should be 
expected to hold in the relativistic quantum context as well.  A quantum 
application of the Postulate 2 condition logically requires a global notion of 
simultaneity. Such a structure already exists {\em a priori} for the non-relativistic case,
but for relativistic applications, seems to require a trajectory ensemble or
related approach.  In any case, KG theory does not provide a simultaneity structure, 
leading to an evident causality paradox.

In previous work,\cite{poirier11nowaveCCP6,poirier12nowaveJCP}
it has been argued that the quantum potential should adhere to a 
{\em universal form}---which is quite independent of the particular 
dynamical law (e.g., non-relativistic vs. relativistic).  This form can 
be derived without reference to the TDSE---although the resultant 
trajectory-based non-relativistic PDE turns out to be equivalent to it, as has been 
mentioned several times.  In the relativistic context, for which there is 
no previously existing PDE with which to compare, the {\em a priori}
 universality of the $Q$ form obviously plays a central role in the 
 development of the formulation.  Finally, of course, the notion of 
 action extremization---so central to classical mechanics, relativity 
 theory and the trajectory-based non-relativistic quantum theory---is
 also applied to the present relativistic quantum application, placing all 
 of these theories within a single, unified framework. 

It is also to be hoped that we may have helped to set the stage
for a possible integration with GR, and with quantum gravity. 
Certainly, our use of general coordinates even in the SR context,
lends itself to this purpose---as does the inherent use of density
fields.  We have,  in any case, already  uncovered some quite intriguing 
similarities---and differences---between the gravitational and 
quantum potentials, as discussed in detail in Sec.~\ref{conversion}. 
Not least of these is the fact that the value of $Q$ itself---and not
 just its spatial gradient---plays a dynamical role in its own right, 
 as seen explicitly in the nontrivial example
of Sec.~\ref{gaussian}.  Also quite relevant (in the non-relativistic context too)
is the fact that very narrow wavepackets experience strongly 
dispersive quantum forces, causing very rapid broadening via
trajectory fanout.  Thus, on small length scales---i.e., on the order of
the Compton wavelength, but still larger than the Planck 
length---quantum trajectory effects may prevent relativistic singularities 
from occurring, or may otherwise play an important dynamical role,
e.g. in inflation. 

There are a number of additional areas for future development
that might be considered.  Some are straightforward, such
as the incorporation of external force fields (e.g., the electromagnetic
field for charged particles),  and a recasting of the various 
density and flux quantities into a stress-energy tensor form (together
with the development of energy and momentum conservation laws).   
Less obvious will be determining how to generalize the present theory
for multiple particles of fixed number---i.e., the relativistic analog
of the many-particle trajectory-based non-relativistic quantum theory 
already in place. Here, the biggest and most important challenge will be 
to define  simultaneity submanifolds for multiple particles.  
This is not so straightforward, and may well prove to be the ``Achilles' heel'' 
of the present approach---most likely related to the oft-quoted claim that relativistic
theories involving interaction energies comparable to particle rest masses must
encompass the creation/annihilation of particles.\cite{bohm,messiah} Of 
course, one might also attempt to apply some of the current ideas in a 
varying-particle-number,  QFT context.  Given the difficulties of the KG equation, 
one wonders whether it is correct to use this as a basis for QFT, as is currently 
accepted practice. How would things change if the present approach
were adopted instead? What impact might this have, e.g., on theoretical
predictions pertaining to the Higgs boson or pi meson? 

Of course, the basic ideas developed here need not be limited to 
massive spin-zero bosons. It is only natural that one might consider
revisiting the Dirac field for spin 1/2 particles, as well as the quantized
electromagnetic field describing photons, armed with the weapons 
developed in the present arsenal.   It is not yet clear what such 
investigations might turn up---i.e., whether they would lead to new
theoretical predictions, as they seem to have done in the present 
case for spin-zero particles, or to predictions in substantive
agreement with currently accepted theories.   In the photon context,
of course, the simultaneity submanifold concept {\em per se} would have 
to be discarded---presumably, to be replaced with an analogous 
family of null hypersurfaces.  It is not yet clear whether this would 
even be possible.

Even for the specific relativistic quantum PDEs developed here, there is still much
work that remains to be done, in terms of characterizing these fully.  Can
an analytical solution for the relativistic Gaussian (or other nontrival) 
wavepacket be derived? Are the solution ensembles always well-behaved 
(subject to the conditions of Sec.~\ref{basic}), or is it possible that 
singularities can somehow develop?  Are the  PDEs inherently nonlinear, 
or can they be recast into an  equivalent linear wave equation? 
Under what conditions is numerical solution of the PDEs unstable, and 
what sorts of algorithms can be used or should be developed?   

One very basic property of particular interest has already been established. 
This is simply that the PDE order in the spatial coordinates  $({\bf C})$ 
is twice that in the time coordinate ($\T$)---even though 
these equations are perfectly invariant with respect to  Lorentz transformations. 
This is not surprising in one sense---i.e., in light of the theoretical
developments presented in this document (notably pertaining to Postulate 2).
In the context of the history of relativistic wave equations, on the other
hand, it is highly unusual---if not borderline heretical.  Traditional treatments
virtually always presume that the space and time orders must be identical,
in order to treat these coordinates on an equal footing. The present 
theory, in contrast, suggests that {\em there is indeed a fundamental difference 
between space and time}---at least when these notions are interpreted in an 
appropriate manner, and linked to a single particle/observer.

In any case, it  should be mentioned that the two-to-one ratio of space-to-time 
orders that characterizes the present relativistic quantum PDEs as described 
above,  also characterizes the non-relativistic quantum PDEs---both the 
trajectory-based PDE, and the TDSE itself. This is, in large measure, why 
the present theory reduces so seamlessly to these others in the non-relativistic limit.   
One might also do well to recall that it is precisely  the {\em lack} of such an asymmetry 
in the KG equation---i.e., the fact that this  equation is second-order in both 
space and time---that is known to lead to the difficulties 
described in Sec.~\ref{KG}, especially that of defining a suitable probability
density quantity.  These problems were well understood in the early days of
the quantum theory---but what is perhaps quite startling is that {\em the same
can be said of all of the theoretical tools used here to develop the present 
fix} (though of course, the trajectory-based formulation itself is new).  In principle, 
Einstein or Dirac, or one of their contemporaries,  could have hit upon this very same 
approach.   One (at least the author) has to wonder how things  
might have progressed  differently, had this idea been born in say, 1932, or even
1952,\cite{bohm52a,bohm52b}  rather than in 2012.

\section*{ACKNOWLEDGEMENTS}

This work was supported by grants from the Robert A. Welch Foundation 
(D-1523) and by a Small Grant for Exploratory Research from the National 
Science Foundation (CHE-0741321).  The author would also like to 
acknowledge useful discussions with Luis Grave de Peralta, George Hinds,  
Jeremy Schiff, and Hung-Ming Tsai.

%
%


\end{document}